\documentclass[longauth]{aa}
\UseRawInputEncoding 
\usepackage[utf8]{inputenc}
\usepackage{pdflscape}
\usepackage{xcolor}
\usepackage{lscape}
\usepackage{float}
\usepackage[hidelinks]{hyperref}
   \hypersetup{
      colorlinks = true,
      citecolor ={cyan}
   }
\usepackage{longtable}
\usepackage{graphicx}
\usepackage{caption} 
\usepackage{txfonts}

\newcommand{\bloem}{BLOeM}
\newcommand{\sky}{{\tt SKY}}

\newcommand{\kms}{km\,${\rm s}^{-1}$}

\newcommand{\gunit}{${\rm cm}\,{\rm s}^{-2}$}

\newcommand{\giraffe}{{\sc giraffe}}

\newcommand{\cms}{\mbox{${\rm cm}\,{\rm s}^{-2}$}}

\newcommand{\Msun}{\mbox{$M_\odot$}}

\def \OeBenum   {82}
\def \Oenum   {20}
\def \Benum   {62}
\def \Onum   {159}
\def \BVnum   {331}
\def \BInum   {303}
\def \Latenum   {136}

\begin{document}

   \title{Binarity at LOw Metallicity (BLOeM)\thanks{Based on observations collected at the European Southern Observatory under ESO program ID 112.25R7.}$^,$\thanks{Table\,\ref{tab:SpT} is available in electronic form at the CDS via anonymous ftp to cdsarc.u-strasbg.fr (130.79.128.5) or via \url{http://cdsweb.u-strasbg.fr/cgi-bin/qcat?J/A+A/}.}} 
    \subtitle{I. A spectroscopic VLT monitoring survey of massive stars in the SMC}

   \author{T.\ Shenar\inst{\ref{inst:TelAv}}
          \and J.\ Bodensteiner\inst{\ref{inst:eso}} 
          \and H.\ Sana\inst{\ref{inst:kul}} 
          \and P.\ A.\ Crowther\inst{\ref{inst:sheffield}} 
          \and D.\ J.\ Lennon\inst{\ref{inst:iac}, \ref{inst:ull}}
          \and M.\ Abdul-Masih\inst{\ref{inst:iac}, \ref{inst:ull}}
          \and L.\ A.\ Almeida\inst{\ref{inst:ufrn}}
          \and F. Backs\inst{\ref{inst:kul}} 
          \and S.\ R.\ Berlanas\inst{\ref{inst:iac}, \ref{inst:ull}}
          \and M.\ Bernini-Peron\inst{\ref{inst:ari}}
          \and J.\ M.\ Bestenlehner\inst{\ref{inst:sheffield}}
          \and D.\ M.\ Bowman\inst{\ref{inst:newcastle}, \ref{inst:kul}}
          \and V.\ A.\ Bronner\inst{\ref{inst:hits}, \ref{inst:uniHD}}
          \and N.\ Britavskiy \inst{\ref{inst:rob}}
          \and A.\ de Koter \inst{\ref{inst:antonpannekoek}, \ref{inst:kul}}  
          \and S.\ E.\ de Mink \inst{\ref{inst:mpa}}
          \and K.\ Deshmukh \inst{\ref{inst:kul}}
          \and C.\ J.\ Evans \inst{\ref{inst:esa_stsci}}
          \and M.\ Fabry \inst{\ref{inst:kul}}
          \and M.\ Gieles \inst{\ref{inst:icrea}, \ref{inst:iccub}}
          \and A.\ Gilkis \inst{\ref{inst:cambridge}}
          \and G.\ Gonz\'alez-Tor\`a\inst{\ref{inst:ari}}
          \and G.\ Gr{\"a}fener \inst{\ref{inst:bonn}}
          \and Y.\ G{\"o}tberg \inst{\ref{inst:ista}}
          \and C.\ Hawcroft \inst{\ref{inst:stsci}}
          \and V.\ H\'enault-Brunet \inst{\ref{inst:smu}}
          \and A.\ Herrero \inst{\ref{inst:iac}, \ref{inst:ull}}
          \and G.\ Holgado\inst{\ref{inst:iac}, \ref{inst:ull}}
          \and S.\ Janssens\inst{\ref{inst:kul}}
          \and C.\ Johnston\inst{\ref{inst:mpa},\ref{inst:kul}}
          \and J.\ Josiek\inst{\ref{inst:ari}}
          \and S.\ Justham\inst{\ref{inst:mpa}}
          \and V.\ M.\ Kalari\inst{\ref{inst:gemini}}
          \and Z.\ Z.\ Katabi\inst{\ref{inst:TelAv}}
          \and Z.\ Keszthelyi\inst{\ref{inst:naoj}}
          \and J.\ Klencki\inst{\ref{inst:eso}} 
          \and J.\ Kub\'at\inst{\ref{inst:ondrejov}}
          \and B.\ Kub\'atov\'a\inst{\ref{inst:ondrejov}}
          \and N.\ Langer\inst{\ref{inst:bonn}}
          \and R.\ R.\ Lefever \inst{\ref{inst:ari}}
          \and B.\ Ludwig\inst{\ref{inst:utoronto}}
          \and J.\ Mackey\inst{\ref{inst:dias}}
          \and L.\ Mahy \inst{\ref{inst:rob}}
          \and J.\ Ma{\'\i}z Apell\'aniz\inst{\ref{inst:cabesac}} 
          \and I.\ Mandel\inst{\ref{inst:monash},\ref{inst:ozgrav}}
          \and G.\ Maravelias\inst{\ref{inst:noa},\ref{inst:forth}} 
          \and P.\ Marchant\inst{\ref{inst:kul}} 
          \and A.\ Menon\inst{\ref{inst:iac}, \ref{inst:ull}} 
          \and F.\ Najarro\inst{\ref{inst:cab}} 
          \and L.\ M.\ Oskinova\inst{\ref{inst:up}}
          \and A.\ J.\ G.\ O'Grady\inst{\ref{inst:carnegie}}
          \and R.\ Ovadia\inst{\ref{inst:TelAv}} 
          \and L.\ R.\ Patrick\inst{\ref{inst:cab}} 
          \and D.\ Pauli\inst{\ref{inst:up}}
          \and M.\ Pawlak\inst{\ref{inst:lund}}
          \and V.\ Ramachandran\inst{\ref{inst:ari}}
          \and M.\ Renzo\inst{\ref{inst:AZ}}
          \and D.\ F. Rocha\inst{\ref{inst:ON_Br}} 
          \and A.\ A.\ C.\ Sander\inst{\ref{inst:ari}}
          \and T.\ Sayada\inst{\ref{inst:TelAv}} 
          \and F.\ R.\ N.\ Schneider\inst{\ref{inst:hits},\ref{inst:ari}}
          \and A.\ Schootemeijer\inst{\ref{inst:bonn}}
          \and E.\ C.\ Sch\"osser\inst{\ref{inst:ari}}
          \and C.\ Sch\"urmann\inst{\ref{inst:bonn}}
          \and K.\ Sen\inst{\ref{inst:umk}}
          \and S.\ Shahaf\inst{\ref{inst:weizmann}}
          \and S.\ Sim\'on-D\'iaz\inst{\ref{inst:iac}, \ref{inst:ull}} 
          \and M.\ Stoop\inst{\ref{inst:antonpannekoek}}
          \and S.\ Toonen\inst{\ref{inst:antonpannekoek}}
          \and F.\ Tramper\inst{\ref{inst:cab}}
          \and J.\ Th. van Loon\inst{\ref{inst:keele}}
          \and R.\ Valli\inst{\ref{inst:mpa}}
          \and L.\ A.\ C.\ van Son\inst{\ref{inst:cca}}
          \and A.\ Vigna-G\'omez\inst{\ref{inst:mpa}}       
          \and J.\ I.\ Villase\~{n}or\inst{\ref{inst:mpia}}
          \and J.\ S. Vink\inst{\ref{inst:armagh}} 
          \and C.\ Wang\inst{\ref{inst:mpa}}
          \and R.\ Willcox\inst{\ref{inst:kul}}
     }

   \institute{
{The School of Physics and Astronomy, Tel Aviv University, Tel Aviv 6997801, Israel\label{inst:TelAv}};\\ \email{tshenar@tau.ac.il}    
   \and
{ESO - European Southern Observatory, Karl-Schwarzschild-Strasse 2, 85748 Garching bei M\"unchen,
Germany \label{inst:eso}}
   \and
{Institute of Astronomy, KU Leuven, Celestijnenlaan 200D, 3001 Leuven, Belgium\label{inst:kul}}
\and
{Department of Physics \& Astronomy, Hounsfield Road, University of Sheffield, Sheffield, S3 7RH, United Kingdom\label{inst:sheffield}} 
    \and
{Instituto de Astrof\'isica de Canarias, C. V\'ia L\'actea, s/n, 38205 La Laguna, Santa Cruz de Tenerife, Spain\label{inst:iac}}
\and    
{Universidad de La Laguna, Dpto. Astrof\'isica, Av.\ Astrof\'sico Francisco S\'anchez, 38206 La Laguna, Santa Cruz de Tenerife, Spain\label{inst:ull}}
\and{Escola de Ciências e Tecnologia, Universidade Federal do Rio Grande do Norte, Natal, RN 59072-970, Brazil}\label{inst:ufrn}
\and
{Zentrum f\"ur Astronomie der Universit\"at Heidelberg, Astronomisches Rechen-Institut, M\"onchhofstr. 12-14, 69120 Heidelberg, Germany\label{inst:ari}} 
\and 
{School of Mathematics, Statistics and Physics, Newcastle University, Newcastle upon Tyne, NE1 7RU, UK\label{inst:newcastle}}
\and
Heidelberger Institut f{\"u}r Theoretische Studien, Schloss-Wolfsbrunnenweg 35, 69118 Heidelberg, Germany\label{inst:hits}
\and
{{Universit\"{a}t Heidelberg, Department of Physics and Astronomy, Im Neuenheimer Feld 226, 69120 Heidelberg, Germany}\label{inst:uniHD}}
\and {Royal Observatory of Belgium, Avenue Circulaire/Ringlaan 3, B-1180 Brussels, Belgium} \label{inst:rob}
\and {{Anton Pannekoek Institute for Astronomy, University of Amsterdam, Science Park 904, 1098 XH Amsterdam, the Netherlands} \label{inst:antonpannekoek}}
\and {Max-Planck-Institute for Astrophysics, Karl-Schwarzschild-Strasse 1, 85748 Garching, Germany\label{inst:mpa}}
\and 
{{European Space Agency (ESA), ESA Office, Space Telescope Science Institute, 3700 San Martin Drive, Baltimore, MD 21218, USA}\label{inst:esa_stsci}}
\and {ICREA, Pg. Llu\'{i}s Companys 23, E08010 Barcelona, Spain}\label{inst:icrea}
\and 
{Institut de Ci\`{e}ncies del Cosmos (ICCUB), Universitat de Barcelona (IEEC-UB), Mart\'{i} Franqu\`{e}s 1, E08028 Barcelona, Spain}\label{inst:iccub}
\and {Institute of Astronomy, University of Cambridge, Madingley Road, Cambridge CB3 0HA, United Kingdom} \label{inst:cambridge}
\and
{Argelander-Institut f\"{u}r Astronomie, Universit\"{a}t Bonn, Auf dem H\"{u}gel 71, 53121 Bonn, Germany\label{inst:bonn}}
\and
{{Institute of Science and Technology Austria (ISTA), Am Campus 1, 3400 Klosterneuburg, Austria}\label{inst:ista}}
\and {Space Telescope Science Institute, 3700 San Martin Drive, Baltimore, MD 21218, USA}\label{inst:stsci}
\and {Department of Astronomy and Physics, Saint Mary's University,
    923 Robie Street, Halifax, B3H 3C3, Canada}\label{inst:smu}
\and {Gemini Observatory/NSF's NOIRLab, Casilla 603, La Serena, Chile}\label{inst:gemini}    
\and
{Center for Computational Astrophysics, Division of Science, National Astronomical Observatory of Japan, 2-21-1, Osawa, Mitaka, Tokyo 181-8588, Japan\label{inst:naoj}}
\and
{Astronomical Institute, Academy of Sciences of the Czech Republic, Fri\v{c}ova 298, CZ-251 65 Ond\v{r}ejov, Czech Republic}\label{inst:ondrejov}
\and
{{Department of Astronomy and Astrophysics, University of Toronto, 50 St. George Street, Toronto, Ontario, M5S 3H4, Canada}\label{inst:utoronto}}
\and{{Dublin Institute for Advanced Studies, DIAS Dunsink Observatory, Dunsink Lane, Dublin 15, Ireland}\label{inst:dias}
}
\and
{Centro de Astrobiolog{\'\i}a (CSIC-INTA), campus ESAC, camino bajo del castillo s/n, 28\,692 Villanueva de la Ca\~nada, Spain\label{inst:cabesac}}
\and
{{School of Physics and Astronomy, Monash University, Clayton VIC 3800, Australia}\label{inst:monash}}
\and
{{ARC Centre of Excellence for Gravitational-wave Discovery (OzGrav), Melbourne, Australia}\label{inst:ozgrav}}
\and 
{IAASARS, National Observatory of Athens, GR-15236, Penteli, Greece}\label{inst:noa}
\and 
{Institute of Astrophysics, FORTH, GR-71110, Heraklion, Greece}\label{inst:forth} 
\and
{Centro de Astrobiolog\'ia (CSIC-INTA), Ctra.\ Torrej\'on a Ajalvir km 4, 28850 Torrej\'on de Ardoz, Spain\label{inst:cab}}
\and
{Institut f\"ur Physik und Astronomie, Universit\"at Potsdam, Karl-Liebknecht-Str. 24/25, 14476 Potsdam, Germany\label{inst:up}}
\and
{McWilliams Center for Cosmology \& Astrophysics, Department of Physics, Carnegie Mellon University, Pittsburgh, PA 15213, USA\label{inst:carnegie}}
\and
{Lund Observatory, Division of Astrophysics, Department of Physics, Lund University, Box 43, SE-221 00, Lund, Sweden}\label{inst:lund}
\and
{{Department of Astronomy \& Steward Observatory, 933 N. Cherry Ave., Tucson, AZ 85721, USA}\label{inst:AZ}
}
\and 
{Observat\'orio Nacional, R. Gen. Jos\'e Cristino, 77 - Vasco da Gama, Rio de Janeiro - RJ, 20921-400, Brazil\label{inst:ON_Br}}
\and
{{Institute of Astronomy, Faculty of Physics, Astronomy and Informatics, Nicolaus Copernicus University, Grudziadzka 5, 87-100 Torun, Poland}\label{inst:umk}
}
\and
{Department of Particle Physics and Astrophysics, Weizmann Institute of Science, Rehovot 7610001, Israel\label{inst:weizmann}
}
\and
{Lennard-Jones Laboratories, Keele University, ST5 5BG, UK\label{inst:keele}}
\and {Center for Computational Astrophysics, Flatiron Institute, New York, NY 10010, USA\label{inst:cca} }
\and
{Max-Planck-Institut f\"{u}r Astronomie, K\"{o}nigstuhl 17, D-69117 Heidelberg, Germany\label{inst:mpia}}
\and 
{Armagh Observatory, College Hill, Armagh, BT61 9DG, Northern Ireland, UK\label{inst:armagh}}
}

   \date{Received -; accepted -}



\abstract{
Surveys in the Milky Way and Large Magellanic Cloud have revealed that the majority of massive stars will interact with companions during their lives. However, knowledge of the binary properties of massive stars at low metallicity, and therefore in  conditions approaching those of the Early Universe, remain sparse. We present the Binarity at LOw Metallicity (BLOeM) campaign, an ESO large programme designed to obtain 25 epochs of spectroscopy  for 929 massive stars in the Small Magellanic Cloud, allowing us to probe multiplicity in  the lowest-metallicity conditions to date ($Z = 0.2\,Z_\odot$).  BLOeM will provide (i) the binary fraction, (ii) the orbital configurations of systems with periods of $P \lesssim 3\,$yr, (iii)  dormant black-hole binary candidates (OB+BH), and (iv)  a legacy database of physical parameters of massive stars at low metallicity.  

Main sequence (OB-type) and evolved (OBAF-type) massive stars are observed with the LR02 setup of the \giraffe\ instrument of the Very Large Telescope (3960 -- 4570\AA\, resolving power $R = 6200$; typical signal-to-noise ratio(S/N) $\approx 70-100$) . This paper utilises the first nine epochs obtained over a  three-month time period.   We describe the survey and data reduction, perform a spectral classification of the stacked spectra, and construct a Hertzsprung-Russell diagram of the sample via spectral-type and photometric calibrations.
 
 Our detailed classification reveals that the sample covers spectral types from O4 to F5, spanning the effective temperature and luminosity ranges $6.5 \lesssim T_{\rm eff} /{\rm kK} \lesssim 45$ and  $3.7 < \log L/L_\odot < 6.1$ and initial masses of $8 \lesssim M_{\rm ini} \lesssim 80\,M_\odot$. The sample comprises \Onum~O-type stars, \BVnum~early B-type (B0--3) dwarfs and giants (luminosity classes V--III), \BInum~early B-type supergiants (II--I), and \Latenum~late-type  BAF supergiants. At least \OeBenum~stars are OBe stars: \Oenum~O-type and  \Benum~B-type (13\% and 11\% of the respective samples).  
In addition, the sample includes 4 high-mass X-ray binaries, 3 stars resembling luminous blue variables, 2 bloated stripped-star candidates, 2 candidate magnetic stars, and 74 eclipsing binaries.

}


   \titlerunning{The Binarity at LOw Metallicity (BLOeM) campaign}
   \authorrunning{T. Shenar et al.}

   \maketitle
%
%
\section{Introduction}\label{sec:intro}

Massive stars ($M_{\rm ini} \gtrsim 8\,M_\odot$), typically classified as OB-type stars on the main sequence, play an increasingly important role in modern astrophysics, with direct links to stellar dynamics, stellar transients, and gravitational-wave (GW) astrophysics \citep{PZ1996, Langer2012, Woosley2017, Wang2020, Marchant2023}. Spectroscopic and interferometric surveys in the Milky Way (MW) and the Large Magellanic Cloud (LMC) have shown that stellar multiplicity is common among massive stars \citep[e.g.][]{Abt1983, Kobulnicky2007, Mason2009, Sana2012, Sana2014, Bordier2024}; over 50\% of massive stars are expected to interact with a companion during their lifetime \citep{Sana2012, Sana2013VFTS, Stegmann2022, Offner2023, Kummer2023}. Such interactions dramatically alter the evolutionary paths and final fates of massive stars, resulting in a plethora of astronomical phenomena such as stripped helium stars and Wolf-Rayet (WR) stars \citep{Paczynski1967, Shenar2016, Goetberg2018, Pauli2022, Drout2023}, rapidly rotating stars with decretion disks \citep[OBe stars;][]{Pols1991, Rivinius2013, deMink2013,  Wang2017, Bodensteiner2020Be, Britavskiy2023, Renzo2021}, stellar mergers and magnetic stars \citep[][]{Ferrario2009, deMink2014, Schneider2019, Shenar2023, Frost2024}, single-degenerate binaries and high-mass X-ray binaries \citep{Corral-Santana2016, Shenar2022BH, Mahy2022}, and GW sources \citep{deMink2016, Marchant2016,  Tauris2017, Mandel2022Rates, Mandel2022Merging}.

Of special interest is massive-star research in low-metallicity (low $Z$) environments, which reflect the conditions prevalent in the distant Universe. An increasing number of transients, such as long-duration $\gamma$-ray bursts \citep[LGRBs;][]{Yoon2005, Woosley2006}, superluminous supernovae \citep[SNe;][]{Quimby2011, Gal-Yam2012}, broad-lined type Ic supernovae \citep{Modjaz2008}, and pair-instability SNe \citep{Barkat1967, Fryer2001, Langer2007, Woosley2017, Farmer2019}, are thought to be associated mainly or exclusively with low-$Z$ conditions. Similarly, the bulk of black-hole (BH) mergers observed with the LIGO-Virgo-KAGRA collaboration are thought to originate in low-metallicity conditions \citep{Abbott2016, Giacobbo2018, Klencki2018}.  

Modern investigations expose deficiencies in our understanding of massive stars at low $Z$. For example, the rates and mass distribution of BH mergers defy original expectations \citep{Broekgaarden2021, Mandel2022Rates, VanSon2022}, and observables such as the rate of LGRBs \citep{Graham2017, Chen2017} and the fraction of OBe stars and Be X-ray binaries \citep{Haberl2016, Schootemeijer2022} as a function of $Z$ are not reproduced by contemporary models \citep{Graham2017, Chen2017, Hastings2021}. Such discrepancies are likely related to insufficient knowledge of massive-star evolution at low $Z$, or to a false extrapolation of the initial conditions of massive stars (e.g. binary fraction, orbital configurations) to low $Z$. The recent detection of a $33\,\Msun$ BH with a low-metallicity companion through {\it Gaia} astrometry provides additional evidence suggesting that low-$Z$ environments are crucial for the formation of massive BHs \citep{Panuzzo2024}.

While it has been shown that the binary properties  of solar-type stars can depend on natal metallicity, this remains a prediction for massive stars \citep{Kroupa2001, Saigo2004, Machida2008, Marks2012, Moe2019, Price-Whelan2020}. 
To mitigate this,
we need spectroscopic monitoring surveys of massive-star populations at different $Z$ environments sensitive to the regime of binary interactions (i.e. orbital periods $0 \lesssim \log_{10}(P\,/{\rm d}) \lesssim 3$). A notable spectroscopic campaign in this context was the VLT-FLAMES Tarantula Survey (VFTS; PI: Evans), which monitored about 1000 massive stars in the Tarantula nebula of the LMC \citep{Evans2011}, which has a metallicity of $\approx 0.5 Z_\odot$. The VFTS survey yielded a comparable intrinsic binary fraction for OB-type stars in the LMC (50-60\%; \citealt{Sana2013VFTS, Dunstall2015}) to that observed in different Galactic environments \citep[50-70\%, e.g.][]{Sana2012, Kiminki2012, Banyard2022, Guo2022}. 
The follow-up Tarantula Massive Binary Monitoring (TMBM; PI: Sana) and B-type Binary Characterisation (BBC; PI: Taylor) programmes also revealed an overall similar distribution of orbital parameters and mass ratios to Galactic samples \citep{Almeida2017, Villasenor2021, Shenar2022TMBM, Mahy2020I}. However, the LMC metallicity only differs by a factor of $\approx 2$ from that of the MW.

The Small Magellanic Cloud (SMC) is a neighbouring dwarf galaxy with $Z \approx 0.2 Z_\odot$ \citep{Hunter2007} located about 62\,kpc from Earth \citep{Graczyk2020}, hosting thousands of massive stars \citep{Humphreys1984}. It had a star-formation peak $10-40\,$Myr ago \citep{Antoniou2010, Rubele2015, Schootemeijer2021}, potentially triggered by a collision with the LMC about 100-150\,Myr ago \citep[e.g.][]{Zivick2018}.  While galaxies of lower metal content exist in the Local Group  (e.g. \object{Sextans A}, \citealt{Skillman1989, Lorenzo2022}; \object{Leo P}, \citealt{McQuinn2015, Evans2019, Telford2023}), the SMC is the only galaxy in which a large sample of massive stars at low $Z$ can currently be resolved and spectroscopically monitored with sufficient spectral resolution and signal-to-noise ratio (S/N). Previous or ongoing surveys addressed aspects related to the SMC massive-star contents  \citep{Humphreys1984, Evans2006, Martayan2007, Evans2004_2dF, Schootemeijer2021},  stellar winds and mass-loss \citep{Ramachandran2019, Vink2023}, and runaway status \citep[RIOTS4,][]{Lamb2016}, but multiplicity has been largely neglected beyond analyses of selected eclipsing binaries \citep{Hilditch2005}, clusters \citep{Dufton2019, Bodensteiner2021}, and individual objects of interest \citep[e.g.][]{Pauli2022AzV}. \citet{Moe2013} investigated the frequency of eclipsing massive binaries among B-type stars in the SMC, LMC, and MW, and found no significant differences between the populations. However, their study was limited to the period range $P \lesssim 20\,$d, and is generally bias-dominated given the small fraction of eclipsing binaries. There is no further information available regarding multiplicity in the SMC.

The need to establish the multiplicity of massive stars at low $Z$ is not the only reason to monitor massive stars in the SMC. Recent spectroscopic monitoring of massive stars in the MW \citep{Mahy2022} and the LMC \citep{Shenar2022TMBM, Shenar2022BH} uncovered the `tip of the iceberg' of a new population of massive single-degenerate binaries: X-ray dormant OB+BH binaries \citep[e.g.][]{Giesers2018}. Such binaries yield precious constraints on core-collapse mechanisms and the presence of SN explosions and possible natal kicks during the collapse into BHs \citep{Mirabel2003, Renzo2019, Atri2019, Langer2020, Banagiri2023}. For example, \citet{VignaGomez2024} recently used the dormant OB+BH binary \object{VFTS 243} to derive a natal kick of 4\,\kms~and an ejection of 0.3\,$M_\odot$ neutrino mass during the collapse of the progenitor. The {\it Gaia} mission will likely uncover dozens more OB+BH binaries in the MW via high-precision astrometry \citep{Mashian2017, Breivik2017, Janssens2022, Janssens2023}, though so far only low-mass stars with BH companions have been discovered \citep{El-Badry2023-GaiaBH1, Chakrabarti2023, El-Badry2023_GaiaBH2, Shahaf2023, Panuzzo2024}. The formation scenarios of these objects are still debated, and it is possible that they originate from dynamical captures, making them less useful for constraints on SN physics \citep{Rastello2023, El-Badry2024, MarinPina2024}. In any case, {\it Gaia} will not uncover extragalactic OB+BH binaries. Finding the first dormant OB+BH binaries in the SMC via spectroscopic monitoring of massive stars has the potential to yield unprecedented constraints on BH formation at low $Z$.

Finally, binary monitoring provides a crucial testbed for single-star models. Eclipsing binaries enable the determination of accurate stellar masses and radii \citep{Hilditch2005, Torres2010, Mahy2020b}. Moreover, well-separated binaries are less likely to have been affected by binary interaction, and hence provide a more solid basis for benchmarking endeavours of single-star models \citep{deMink2014}.

Motivated by these objectives, we initiated  a novel spectroscopic monitoring survey of a large population of  massive stars in the SMC. The Binarity at LOw Metallicity (BLOeM) campaign is a European Southern Observatory (ESO) Large Programme (PI: Shenar, dPI: Bodensteiner; ID: 112.25R7) scheduled for 2023 -- 2025. Relying on 116\,hr of observing time with the Fibre Large Array Multi Element Spectrograph (FLAMES; \citealt{Pasquini2002}) of the Very Large Telescope (VLT), the survey is underway and assembling 25 epochs of spectroscopy for 929 massive stars for a total baseline of two years (four semesters). The survey will enable a full characterisation of the binary fraction and the orbital parameters of stars with orbital periods of up to a few years and with mass ratios of as low as $M_2/M_1 \approx 1/10$; the discovery of dormant OB+BH binaries; and a complete analysis of the binary and single-star content of the sample. 

This first paper in the series  provides an overview of the sample and sample selection (Sect.\,\ref{sec:sample}), a description of the data reduction and quality (Sect.\,\ref{sec:data}), a detailed spectral classification (Sect.\,\ref{sec:SpT}), and a first characterisation of the physical mass range and evolutionary status of the sample stars (Sect.\,\ref{sec:results}), followed by our main conclusions (Sect.\,\ref{sec:conclusions}).


\section{Sample selection}\label{sec:sample}

\begin{figure}
\centering
\includegraphics[width=0.5\textwidth]{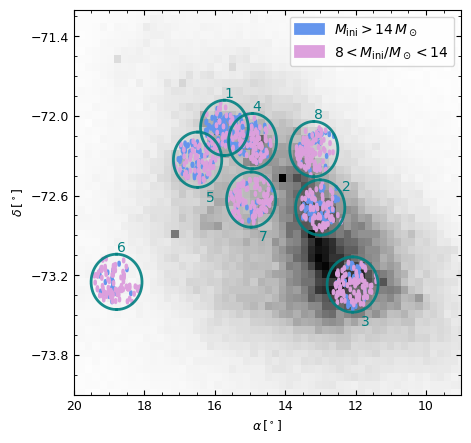}
\caption{The eight FLAMES pointings marked on a density map of the underlying {\it Gaia} source catalogue of the SMC (with $G < 19\,$mag) as a function of right ascension ($\alpha$) and declination ($\delta$) (darkest pixels correspond to $\approx 900$ stars). The green rings correspond to the FLAMES FoVs, which are 25' in diameter. The 929 targets are shown as blue and pink dots based on their estimated initial masses (see legend and text). We note that the regions most densely populated with stars in the SMC (e.g. the bar) are not rich in massive stars, and hence only a few fields were allocated there. }
\label{fig:SMCDensity}
\end{figure}

The BLOeM sample (see Fig.\,\ref{fig:SMCDensity}) was selected from the third {\it Gaia} data release catalogue \citep[{\it Gaia} DR3,][]{Gaia2023DR3}. The catalogue was retrieved from the {\it Gaia} database using a search radius of $2.6^\circ$ around the SMC centre ($\alpha[{\rm hrs]}, \delta[\rm deg]$ = 00:52:38, -72:48:01; epoch J2000). To achieve S/N $\gtrsim 20$ per pixel (0.2\,\AA~spectral bin), only stars with $G < 16.5\,$mag were retrieved. Foreground objects were filtered via two constraints. First, the parallax $\pi$ was required to be consistent with zero\footnote{{\it Gaia} is not sensitive enough to measure non-vanishing parallaxes within errors at the SMC distance.} within 5$\sigma$, that is $\pi/\sigma_\pi < 5$. Second, the  proper motions of the stars were required to be consistent within 15$\sigma$ with the SMC proper motion ($\mu_\alpha = 0.695\pm0.240\,{\rm mas}\,{\rm yr}^{-1}$ and $\mu_\delta = -1.206\pm0.140\,{\rm mas}\,{\rm yr}^{-1}$) following \citet{Yang2019} and \citet{Schootemeijer2021}.

\begin{figure*}
\centering
\includegraphics[width=\textwidth]{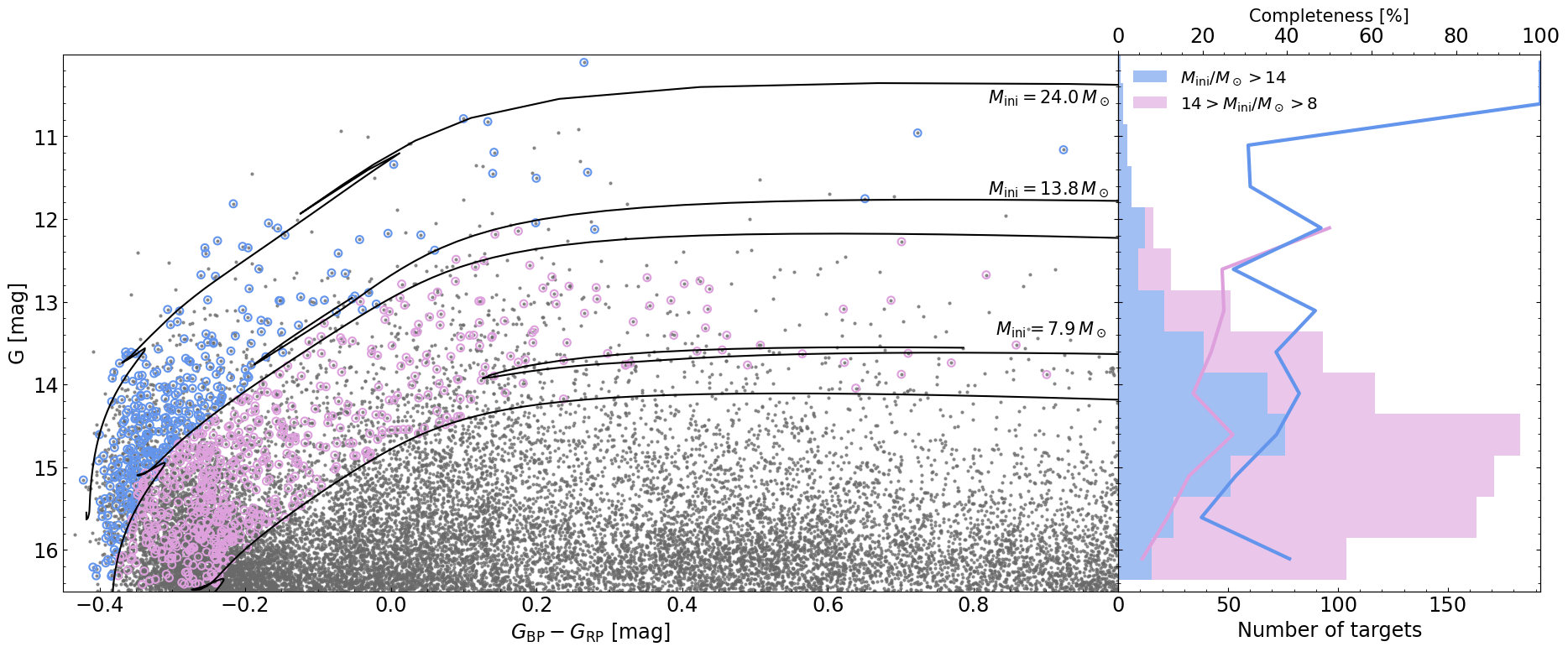}
\caption{Completeness of the BLOeM dataset with respect to the underlying {\it Gaia} catalogue. {\it Left:} CMD showing the underlying {\it Gaia} catalogue used to choose the BLOeM sample (black dots). Three evolution tracks computed by \citet{Schootemeijer2019} and \citet{Hastings2021} with the MESA stellar evolution code (see Sect.\,\ref{subsec:HRD}) for $M_{\rm ini} = 7.9, 13.8$, and $24\,M_\odot$ are plotted. The tracks were adjusted to the SMC distance and an average reddening and extinction (see text for details). BLOeM targets are encircled with blue ($M_{\rm ini}/M_\odot \gtrsim 14$; born as O-type stars) and pink ($8 \lesssim M_{\rm ini}/M_\odot \lesssim 14$; born as B-type stars) circles. {\it Right:} Magnitude distribution of the subsamples, along with completeness fractions for the two subsamples with respect to the underlying {\it Gaia} SMC catalogue.}
\label{fig:BLOeMOverview}
\end{figure*}

We omitted the 12 known SMC Wolf-Rayet (WR) stars \citep{Neugent2018} from the sample, because they have been previously monitored \citep{Foellmi2003, Hainich2015, Shenar2016, Schootemeijer2024}. Finally, we omitted potential red supergiants (RSGs) from the sample by imposing $G_{\rm BP} - G_{\rm RP} < 1\,$mag: due to the large radii of RSGs, RSG binaries have periods that exceed a few years \cite[e.g.][]{Patrick2019, Neugent2020} and hence exceed the two-year baseline of our programme. 

To avoid crowding, we removed objects that have a {\it Gaia} source brighter than $G=19\,$mag closer than 1.2'', which corresponds to the FLAMES fibre size. We also explicitly excluded stars within 30'' of the centres of the dense SMC clusters \object{NGC\,330} and \object{NGC\,346}.

we made use of an evolutionary track of a $M_{\rm ini} = 8\,M_\odot$ star (see Sect\,\ref{subsec:HRD}) computed by \citet{Schootemeijer2019} and \citet{Hastings2021} with

In an attempt to select only massive stars, we made use of an evolutionary track of a $M_{\rm ini} = 8\,M_\odot$ star (see Sect.\,\ref{subsec:HRD}) computed by \citet{Schootemeijer2019} and \citet{Hastings2021} with the Modules for Experiments in Stellar Astrophysics (MESA) stellar evolution code \citep{Paxton2011, Paxton2013, Paxton2015, Paxton2018, Paxton2019, Jermyn2023}. We converted the physical parameters along the track to a $G$-band magnitude and $G_{\rm BP} - G_{\rm RP}$ colours using bolometric corrections taken from the MIST webpage\footnote{\url{http://waps.cfa.harvard.edu/MIST/model_grids.html}; the bolometric corrections were retrieved by fitting a fifth-order  polynomial to MIST values for $\log g = 3\,$\cms~and $Z = 0.18\,Z_\odot$.} 
\citep{Dotter2016, Choi2016}. We adjusted the track on the colour--magnitude diagram (CMD) by adopting a distance of 62\,kpc \citep{Graczyk2020}, and assuming an average value for the reddening of $E_{\rm BP - RP} = 0.14\,$mag and extinction of $A_G = 0.28\,$mag \citep{Schootemeijer2021}.  We then only selected targets whose CMD positions lie above this track before becoming a RSG (effective temperature $T_{\rm eff} > 6\,$kK; see Fig.\,\ref{fig:BLOeMOverview}). This resulted in a massive-star catalogue of 5576 stars subject to the criteria above. We also made use of a MESA track computed for $M_{\rm ini} = 14\,M_\odot$ by \citet{Schootemeijer2019} to divide the sample into stars with $M_{\rm ini} \gtrsim 14\,M_\odot$ (born as O-type stars) and $M_{\rm ini} \lesssim 14\,M_\odot$ (born as B-type stars)\footnote{The threshold mass for O-type stars is typically taken as $15\,M_\odot$ for the Galaxy \citep[e.g.][]{Martins2005}. However, stars in the SMC are more compact and hot at a fixed mass \citep[e.g.][]{Georgy2013}, such that this threshold is likely lower at low $Z$.}. While massive stars are typically born as OB-type on the main sequence, they can appear as OBAF blue/yellow supergiants after leaving the main sequence (in addition to Wolf-Rayet stars and GMK red supergiants, which were omitted from our survey, as described above). Hence, we can expect the spectral types of the sample stars to span the entire OBAF range.

Our programme includes a total of eight FLAMES plate configurations, each with a field-of-view (FoV) of 25' in  diameter, although the instrument setup ensures visibility of targets only within a 20' diameter (Fig.\,\ref{fig:SMCDensity}). For each FLAMES plate configuration, 130 \giraffe\ fibres are available. We allocated 14 fibres for sky, leaving each field with 116 science targets. The only exception is field 8, for which 13 sky fibres and 117 science targets are available.  This makes a total of $7 \times 116 + 117 = 929$ science targets. 

To obtain a balanced sample of 929 science targets out of the 5576 available targets, we aimed to achieve a $G-$band magnitude distribution that is as homogeneous as possible, while prioritising the brightest and hence most massive stars, which are rarer.  The target allocation was then followed via the two steps described below. 

First, the centre of a FLAMES pointing was chosen.  The choice of field centre followed automatically by identifying the coordinate that encloses as many massive and bright stars as possible within a circle of  25' in diameter centred on that coordinate, after removing stars which were already allocated in previous field allocations. Specifically, the pointings were selected by identifying the centre coordinates that result in the largest number of stars with $M_{\rm ini} \gtrsim 14\,M_\odot$ (see above) and $G < 14.7\,$mag. This resulted in fields 1 -- 8 shown in Fig.\,\ref{fig:SMCDensity}, which provide a good coverage of the massive-star content of the SMC. We note that while some of the fields overlap (e.g. 1 and 4), there is no overlap between the allocated stars within each field.

As a second step, for each plate configuration, targets within a circle of  20' in diameter were allocated to the available fibres, starting from the brightest ones, while aiming to achieve a homogeneous sampling across the $G$-band. Specifically, we divided the magnitude range 10 -- 16.5\,mag into 15 bins, and aimed to achieve homogeneous coverage across these bins, resulting in $7 - 8$ stars per magnitude bin, per field. This was not always possible, given the rarity of bright stars, and the smaller parameter range of massive stars at lower magnitudes. As the final fibre allocation depends on limitations related to the FLAMES fibre positioner, the remaining massive stars in each field within a circle of 25' in diameter were taken as backup targets, with their priority sorted by brightness.

The final fibre allocation was then performed using ESO's Fibre Positioner Observation Support Software (FPOSS). The majority of our input targets made it to the final allocation, but the final sample includes a few  dozen backup targets. Sky fibres were allocated to 14 fibres in each field (13 for field 8) selected from concentric rings around the field centre where no known {\it Gaia} source is located. Finally, guide stars were selected following requirements in the ESO FLAMES manual for cycle P112. 

The final sample of 929 targets is shown on a CMD in Fig.\,\ref{fig:BLOeMOverview}, along with a magnitude histogram. We also show the completeness fraction with respect to the underlying SMC {\it Gaia} catalogue as a function of $G$-band magnitude.  Of the 929 stars, and based on the single-star tracks in Fig.\,\ref{fig:BLOeMOverview}, 323 have initial masses of above $\approx 14\,M_\odot$  ("born as O-type stars") and 606 are below this mass (`born as B-type stars'). Evidently, the sample reaches a completeness fraction of $\gtrsim 40\%$ for the $M_{\rm ini} \gtrsim 14\,M_\odot$ subsample, and $\gtrsim 20\%$ for the $ 8\,M_\odot \lesssim M_{\rm ini} \lesssim 14\,M_\odot$ sample. 

The naming convention for the sample stars follows the format F-NNN, where F is the field number (1 -- 8), and NNN is the target number (001 -- 117), sorted by ascending right ascension per field.

\section{Observations and data reduction}\label{sec:data}

At the time of writing, 9 out of 25 epochs were obtained during the first semester and were processed in the framework of the BLOeM survey. The field centres, along with the MJD values of the epochs acquired so far, are provided in Table\,\ref{tab:MJDs} of Appendix\,\ref{sec:EpochSpT}.

The data reduction was performed with the \giraffe\ pipeline v.\,2.16.11 under the ESO CPL environment (v.\,3.13.5). Each exposure was split into two subexposures for a robust removal of cosmic rays (cosmics, Sect.\,\ref{subsec:coscor}). The data reduction itself consisted of four steps: bias and dark subtraction, flatfield correction, and wavelength calibration. All spectra were resampled by the ESO CPL pipeline to a constant wavelength step of 0.2\AA\ (see also Sect.~\ref{subsec:wlc}) and science spectra were then sky-subtracted and corrected for the barycentric motion. As a final step, we resampled the individual spectra to a common wavelength grid and co-added the spectra of individual targets to boost S/N in order to  improve spectral typing. 
The spectra have a spectral resolving power of $R \approx 6200$ and cover the spectral range $3960 - 4570\,$\AA, with a median S/N of $70 - 100$ per pixel and epoch; details are provided below. We provide further details below on specific aspects of the data reduction process\footnote{The reduced data and co-added spectra will be made available via ESO phase 3 upon termination of propriety time; they are currently available on \url{http://www.astro.tau.ac.il/~tshenar/DR3/}; please contact T.\ Shenar or J.\ Bodensteiner for credentials. }.

\subsection{Temporal sampling}\label{subsec:TempSamp}

The temporal sampling of the epochs is not strictly defined a priori in order to allow for sufficient scheduling flexibility. We insist on a minimum separation between each epoch of 1\,d, and a maximum of 20\,d to ensure that all epochs are acquired within a semester. The typical separations between the epochs used here, acquired during September\ 2023 through December\ 2023, are days to weeks (see Table\,\ref{tab:MJDs}), for a total time baseline of 50 -- 70\,d, depending on the field. The fact that the acquisition will take place across four semesters ensures that both short-scale and long-scale variability will be covered by the survey. As a multiplicity analysis is beyond the scope of the present paper, we refrain from a complete characterisation and Fourier mapping of the temporal sampling here.

\subsection{Wavelength calibration}\label{subsec:wlc}

 We paid particular attention to the quality of the wavelength solution. For FLAMES \giraffe, a reference ThAr calibration frame in the LR02 setup is obtained at the end of an observing night by illuminating each fibre on a given plate with the light of a ThAr lamp. As a result, each fibre of each epoch has its own calibration ThAr spectrum, which is used by the CPL pipeline to produce a 2D polynomial dispersion solution. The wavelength calibration solution was performed in two steps. First we used the instrument model provided by the pipeline static calibration v2.16.11 to compute a first-guess solution and run a first iteration of the {\tt giwavecalibration} CPL recipe. We modified the standard options to allow for a large detection window of 20 pixels at first, for five iterations, and then progressively reduced it to 15 and ultimately 10 in the remaining five iterations. We decreased the rejection threshold from 1.2 to 3$\sigma$, allowing us to retain a greater number of lines and provide a first wavelength solution with a root mean square (rms) residual of 0.45-0.56 pixels. By comparing ThAr spectra of different nights we noticed a  slight drift in the wavelength solution in various parts of the ThAr spectrum, with a higher stability in the centre of the wavelength range and a larger epoch-to-epoch variation in the blue and red parts of the wavelength solution. We estimated the internal consistency to be no better than a few km\,s$^{-1}$. To improve on this,  we performed a second iteration using the first solution as a new input guess-solution and reiterating the {\tt giwavecalibration}  recipe. This yields a final dispersion solution, characterised with a rms residual in the range of 0.22 to 0.25 pixels. 

Three epochs of field 1 were observed with the SIMCAL lamp on at the beginning of the survey. The SIMCAL lamp yields a set of five ThAr spectra spread across the detector and acquired simultaneously with the science observations. This setup was discontinued because the glow of the strongest ThAr lines impacted the  signals of the adjacent fibres, leaving a noticeable imprint of ThAr on nearby sky and weak objects. Yet, we noticed no difference in the quality of the wavelength solution with or without the SIMCAL lamp.

We experimented with the pipeline rebin pixel size ({\tt --rbin-lstep}) but did not find this to yield any significant improvement and we decided to resort to the default resampling of 0.2\,\AA. The average spectral resolving power measured on the ThAr lines across all epochs is $R = 6224 \pm 90$.

Finally, we investigated the stability and consistency of the individual dispersion solutions across the \bloem\ dataset. We specifically investigated the inter-epoch stability for given fibres as well as the intra-epoch consistency across all fibres of a given epoch. For the first one, which informs us about the temporal stability of the data, we cross-correlated the  wavelength-calibrated ThAr spectra of a given fibre and field with that of the same fibre and field across all the epochs obtained so far. We found  maximum peak-to-peak differences to be of 130~m\,s$^{-1}$, with a standard deviation  of below 50~m\,s$^{-1}$. For the latter, which informs us about the consistency of the dispersion solution across the \bloem\ data sets, we cross-correlated the ThAr spectra of all fibres for a given epoch and field with one arbitrary ThAr spectrum (fibre \#10) for the epoch under consideration. While the peak-to-peak and rms variations are slightly larger (750 and 150~m\,s$^{-1}$ in the worse case), they remain well within the specifications of the instrument.  For a few nights, no ThAr calibration frame could be obtained in the morning following the observations. In such cases, we use the  frame closest in time, typically the one from the morning before. However, in eight cases we had to resort to ThAr calibration frames taken 30 to 40~h before of after observations of our targets. Nevertheless, no noticeable difference in the quality of the calibration could be found, again suggesting that temporal drifts are limited.

\subsection{Sky subtraction}\label{subsec:sky}

Sky spectra were obtained simultaneously with the science integration through a set of fibres allocated to empty patches of sky. These are dubbed \sky\ fibres and record all background signal, including the moon and nebular and sky emission spectra depending on the wavelength regime. Hence, not all signal recorded by \sky\ fibres is from the `sky' itself, but we nonetheless adopt the generic denomination here.

As described earlier, we typically used 14 \sky\ fibres in each field and kept the location constant across all epochs of a given field. We visually inspected the \sky\ spectra from each field and each epoch and flagged spectra that seemed  to be significantly higher than the median of the epoch and field. These are possibly contaminated by faint objects and therefore not representative of the true background signal. Once a sky location has been flagged as contaminated in any of the epochs of a given field, it is rejected from all epochs so that the median sky is always computed with the same set of input locations. In this process, we rejected 4, 3, 1, and 2 sky fibres for fields 1, 2, 5, and 6, respectively. The sky correction was finally performed by subtracting the median spectra of the `valid' \sky\ fibres at the corresponding field and epoch. The error sky spectrum was computed as the rms around the median sky at each pixel after masking sky pixels affected by cosmic rays through a $\kappa-\sigma$ iterative filtering. The error sky spectrum was added quadratically to the error science non-sky-subtracted spectrum produced by the pipeline in order to compute the  error spectrum of the sky-subtracted science data.

Of importance, any nebular component in the sky spectra is obtained at the position of the \sky\ fibres. In our adopted observational setup, these are located tens of arcsecs to several arcminutes away from any science spectra and hence they do not reflect the local nebular conditions of any science targets. The process of taking the median across 10 to 14 sky positions ensures that any local nebulosity in the \sky\ fibres is averaged out.  As a corollary, the sky-subtracted science spectra are not corrected for nebulosity and therefore retain their nebular component.

\subsection{Normalisation}
\label{subsec:Norm}

The sky-subtracted spectra of each subexposure were individually and automatically normalised using a designated Python script to remove the underlying continuum, which is a combination of the stellar continuum, reddening, and instrumental response.  Overall, the non-normalised spectra are well-behaved and have a smooth, typically monotonically decreasing behaviour across the spectral range. The normalisation was performed by fitting a polynomial to automatically identified continuum points along the spectrum. Following several independent attempts, we found that a polynomial of degree eight provided the best results in terms of robust normalisation; lower polynomial degrees resulted at times in underfit regions, while higher degrees introduced wavy patterns in the spectra, which can impact subsequent analysis.

The continuum points were automatically selected in an iterative manner. For this purpose, we computed an average spectrum with a small 3-pixel window and a median spectrum with a large 250-pixel window, which is a first approximation for the continuum. A first batch of continuum points was defined as the set of points whose average flux is 2$\sigma$ above the median flux, with $\sigma$ originating in the error spectrum. To ensure that the edges of the spectra are included, we always include the first and last 10 pixels not affected by cosmics in this set. We then fit a polynomial of degree eight through these points to obtain an approximation for the continuum. This process is then repeated three times, with the polynomial fit for the continuum replacing the role of the median spectrum of the original iteration. 

\subsection{Cosmic correction and combination of subexposures}
\label{subsec:coscor}

\begin{figure}[htp]
\centering
\includegraphics[width=.5\textwidth]{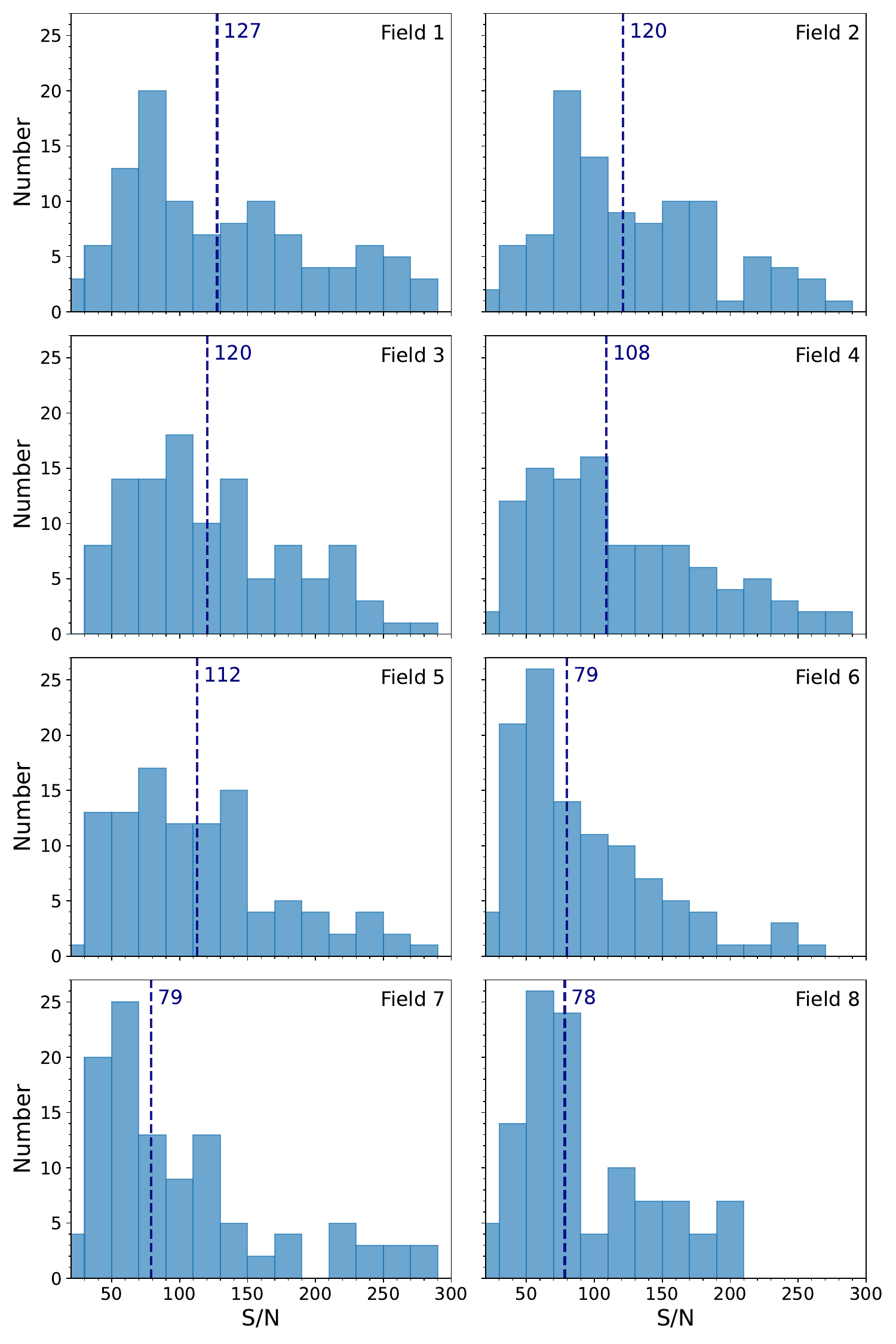}
\caption{Histograms of the median S/N per physical 0.2\,\AA~pixel across the nine available epochs, for each of the eight fields of the sample. Red dashed lines and labels denote the median of all median S/N values per field. }
\label{fig:SNR}
\end{figure}

To boost the S/N, the two subexposures of each epoch and star were combined into one exposure. Moreover, we used the availability of two subexposures for a robust cosmic correction using a designated Python script. The process is as follows. First, positive flux outliers are identified in each of the two subexposures via the condition \mbox{$f_{\rm 1,2}(\lambda) >  {\rm min}\left\{f_1(\lambda), f_2(\lambda)  \right\} + 6\sigma$}, where $\sigma(\lambda)$ is the minimum of the error spectra of both subexposures. Cosmics are identified as points  flagged as such in one spectrum but not in the other -- this ensures that intrinsic emission features (e.g. disc features, wind features, nebular lines) are not removed via this process. In  principle, cosmics may be present in the same pixels in both subexposures, but such cases occur so rarely that this does not warrant further consideration. The cosmics are then removed by replacing each pixel identified as cosmic in one subexposure with the flux value of the same pixel in the second subexposure. The flux of the remaining pixels is formed by quadratically adding the two subexposures. The S/N per pixel ranges between 20 and 300, with a median value of between 70 and 120, depending on the field (Fig.\,\ref{fig:SNR}).

\subsection{Co-added spectra}
\label{subsec:coadd}

\begin{figure*}[htp]
\centering
\includegraphics[width=\textwidth]{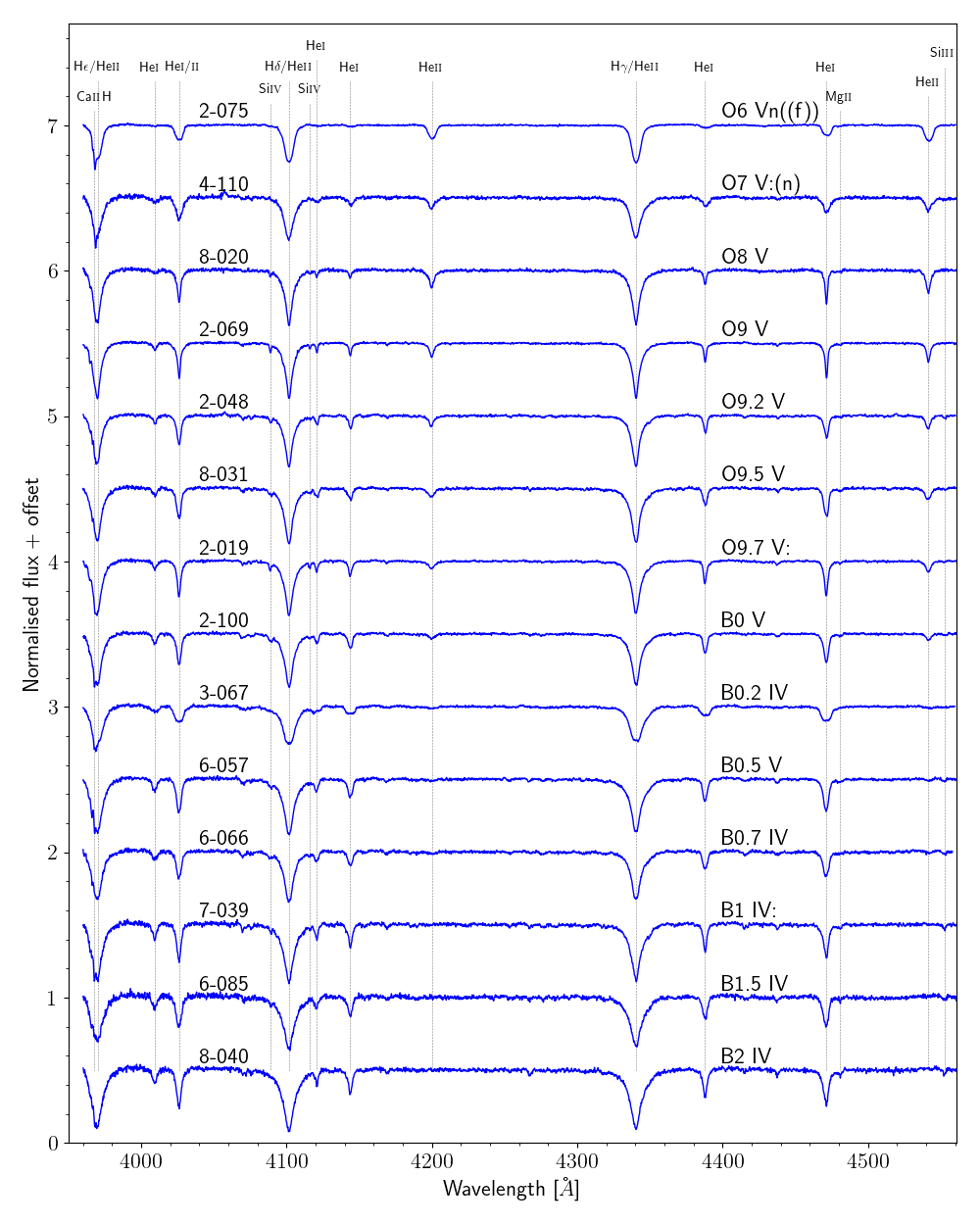}
\caption{Montage of the normalised co-added spectra formed from the nine available epochs for selected OB stars classified as dwarfs (luminosity class V) or sub-giants (IV). The spectra are shifted by a constant for clarity.  BLOeM IDs and spectral types are noted above each spectrum, and diagnostic lines are identified.}
\label{fig:Montage1}
\end{figure*}

\begin{figure*}[htp]
\centering
\includegraphics[width=\textwidth]{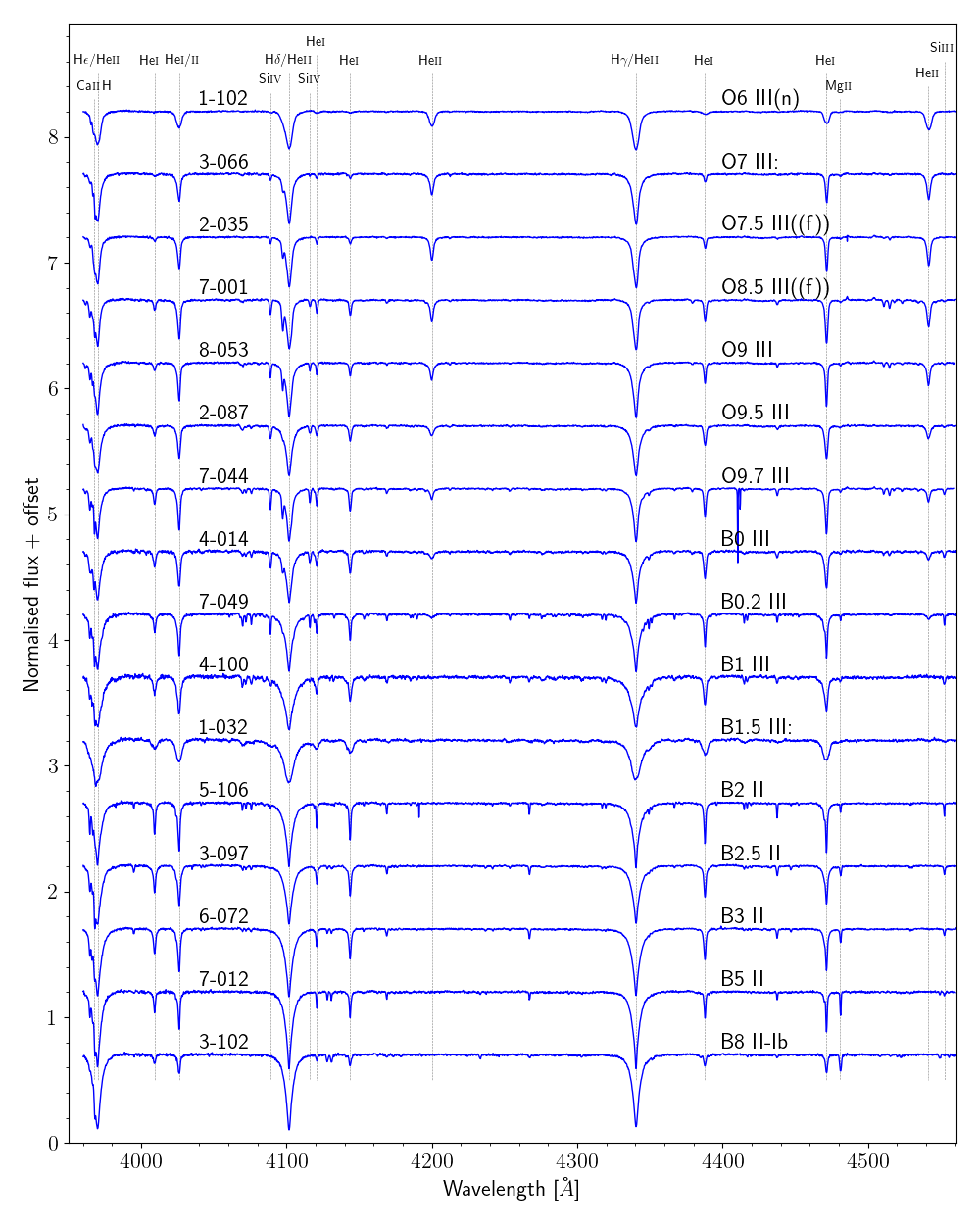}
\caption{Montage of a selection of giant (luminosity class III) and bright giant (II) OB stars in the BLOeM sample, ordered from early (top) to late (bottom) type. BLOeM IDs and spectral types are noted above each spectrum. Diagnostic lines are identified.}
\label{fig:Montage2}
\end{figure*}

\begin{figure*}[htp]
\centering
\includegraphics[width=\textwidth]{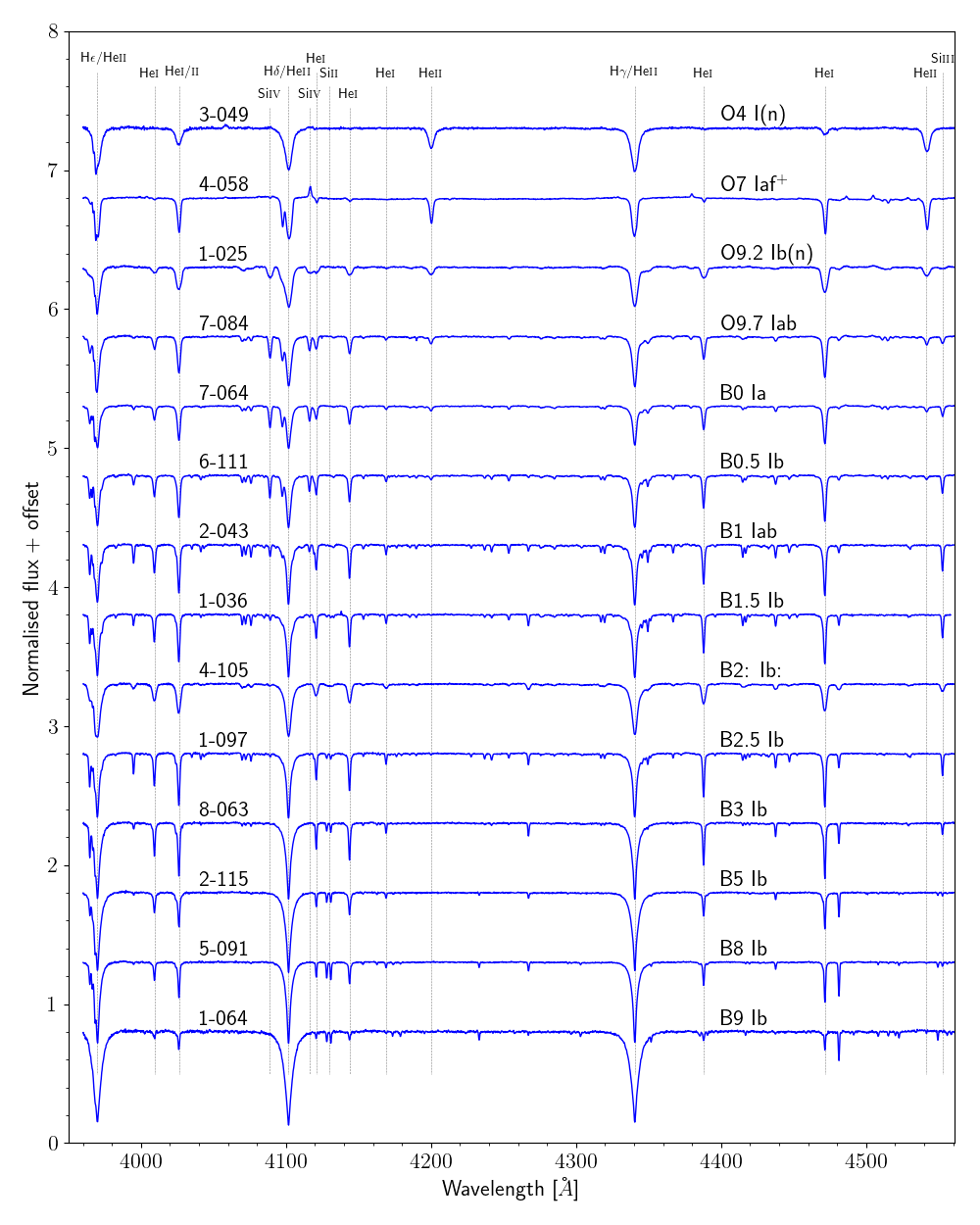}
\caption{Montage of a selection of supergiant OB stars in the BLOeM sample, ordered from early (top) to late (bottom) type. BLOeM IDs and spectral types are noted above each spectrum. Diagnostic lines are identified.}
\label{fig:Montage3}
\end{figure*}

\begin{figure*}[htp]
\centering
\includegraphics[width=\textwidth]{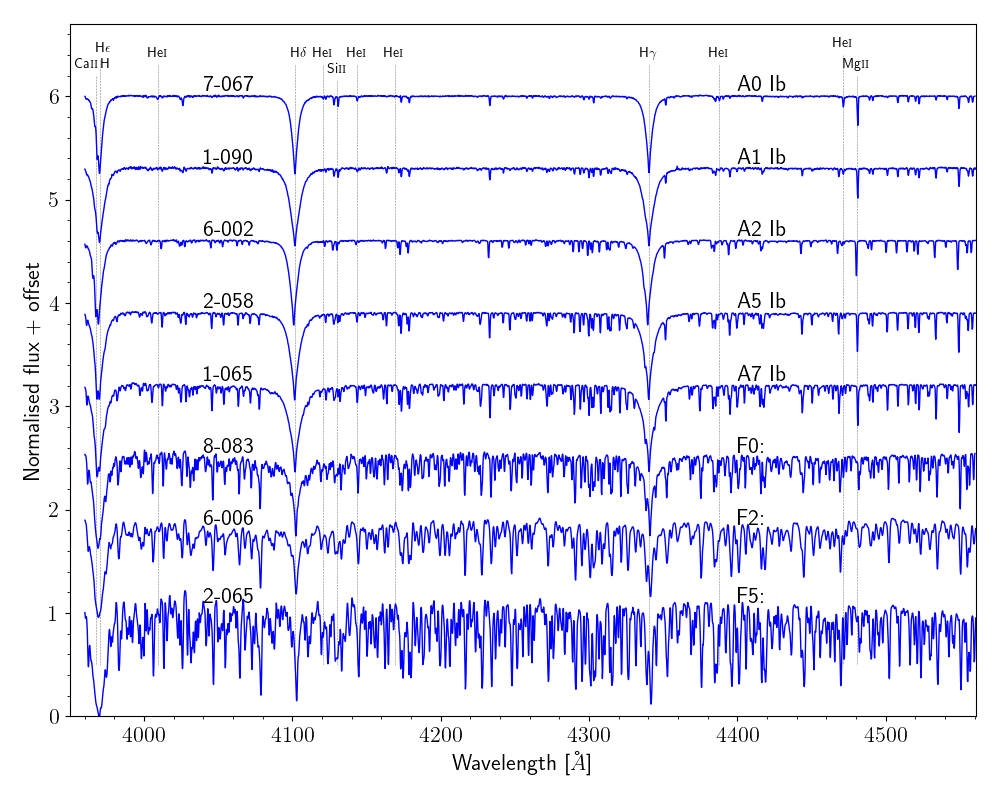}
\caption{Montage of a selection of AF supergiants in the BLOeM sample, ordered from early (top) to late (bottom) type. BLOeM IDs and spectral types are noted above each spectrum. Diagnostic lines are identified.}
\label{fig:Montage4}
\end{figure*}


For a more robust spectral classification, we produced high-S/N spectra by stacking the nine available epochs for each target. To achieve this, the cross-correlation occurred in two steps. First, the spectra were shifted to the rest frame by cross-correlating them with a spectral model and then co-added, which enabled us to generate a first co-added spectrum. In this step, the entire spectral range was considered, which is dominated by the Balmer lines H$\delta$ and H$\gamma$. As a second step, we cross-correlated all the epochs against the co-added spectrum formed in the previous step, which is calibrated to the rest frame. However, this time we used a narrow spectral window around  He\,{\sc i}\,$\lambda 4471$ (namely 4460--4485\,\AA), which is present in almost all objects\footnote{For AF supergiants, this range is rich in other spectral lines, such as Mg\,{\sc ii}\,$\lambda 4481$, making the cross-correlation in this range equally possible.}. This allows a more accurate radial-velocity (RV) measurement that is tuned to the spectral morphology of the individual star. For the first step, we used a precomputed model from the TLUSTY O-star model atmosphere grid \citep{Hubeny1995, Lanz2003} for $Z = 0.1\,Z_\odot$\footnote{This is the closest available metallicity to that of the SMC. In any case, the impact of the metallicity on RV measurements is negligible}, an effective temperature of $T_{\rm eff} = 30\,$kK, and a surface gravity of $\log g = 4.0\,$[\gunit]. The exact choice of the model can impact the absolute RV calibration, but has no impact on the process of spectral classification, which is the focus of this paper.

We note that for double-lined spectroscopic binaries  (SB2), the process of co-adding the data will smear the spectral features of the two components and result in an average spectrum, which may well contaminate the classification.  Clear cases of SB2s (40/929 in total)  were identified via visual inspection of the available epochs (see also Sect.\,\ref{sec:SpT}). A montage of the co-added spectra of selected stars ordered by descending spectral type (Sect.\,\ref{sec:SpT}) is shown in Figs.\,\ref{fig:Montage1} -- \ref{fig:Montage4}.

\section{Spectral classification}\label{sec:SpT}

We established SMC reference OB stars from comparison with Galactic O star templates from \citet{Sota2011} and \citet{Maiz2016} or Galactic B star templates from \cite{Negueruela2024}. These stars are drawn from BLOeM datasets, supplemented by archival VLT/FLAMES \citep{Evans2006, Dufton2019} or VLT/X-Shooter \citep{Vink2023} datasets. Reference stars are ideally sharp-lined, permitting rotational broadening to be applied for comparison with fast rotators. We assign e, e+, and pe for stars exhibiting (i) H\,{\sc i}, (ii) H\,{\sc i} and Fe\,{\sc ii}, and (iii) H\,{\sc i} and  He\,{\sc i} emission, respectively. Objects whose spectra are contaminated by nebular emission are designated `neb'. In some cases, it was difficult to discern between intrinsic source emission and nebular contamination. To  distinguish between these cases, we made use of images acquired with the wide-field Digitized Sky Survey (DSS) 2-red to identify objects that are embedded within nebulous regions. More details are given in Appendix\,\ref{sec:DSS}.  The qualifier `:' deems the classification uncertain. We note that the BLOeM spectral range includes most diagnostic lines, but misses a few lines that can help refine the classification, such as N\,{\sc iii}\,$\lambda \lambda 4634, 4042$, He\,{\sc ii}\,$\lambda 4686$, H$\beta$, and H$\alpha$.

Spectral types of O stars are determined following the Galactic O star scheme of \citet{Sota2011} and \citet{Maiz2016}, which utilises the ratio of He\,{\sc ii} $\lambda$4542 to He\,{\sc i} $\lambda$4471 or $\lambda$4388 for late subtypes (O8.5+). Luminosity classes are usually assigned courtesy of He\,{\sc ii} $\lambda$4686. As this line is not available for the BLOeM dataset, luminosity classes for O4--7 stars are estimated from the detection of N\,{\sc iv} $\lambda$4058 and Si\,{\sc iv} $\lambda\lambda$4088-4116 emission lines. For late O subtypes, we use a combination of H$\gamma$ and H$\delta$ line profile morphologies supplemented by the ratio of Si\,{\sc iv}\,$\lambda$4088, $\lambda$4116 to He\,{\sc i} $\lambda$4121, $\lambda$4144 lines, following \citet{Walborn1990}. It is well known that incorrect luminosity classes would be assigned for silicon-poor SMC stars based on Galactic templates \citep{Walborn1983}. For example, \citet{Walborn2014} showed that using Si-He criteria results in luminosity classes that are 
different from those resulting from the use of He-line criteria alone. Hence, SMC reference stars are essential for this approach.  Nomenclature linked to the region of He\,{\sc ii} $\lambda$4686 ---such as f and (f), which reflect the presence of emission in the line--- is not possible, but we are able to flag potential rapidly rotating O stars via (n), n, nn, and so on, on the basis of the FWHM of He\,{\sc ii} $\lambda$4542 (early and mid O-types) and He\,{\sc i} $\lambda$4471 (late O-types). For a few O-type stars, spectral types (mostly luminosity classes) were adjusted based on previous literature, given the lack of diagnostics such as He\,{\sc ii}\,$\lambda 4686$ and H$\beta$ in the BLOeM dataset. These cases are documented in the comments in Table\,\ref{tab:SpT}.

We follow the SMC B-type scheme of \citet{Lennon1997}, supplemented by early B templates from \cite{Negueruela2024}, in assigning spectral types from Si\,{\sc iv} $\lambda$4088, Si\,{\sc iii} $\lambda$4553, and Si\,{\sc ii} $\lambda\lambda$4128--32, plus the ratio of He\,{\sc i} $\lambda$4471 to Mg\,{\sc ii} $\lambda$4481 at late subtypes. Secondary criteria for B0--0.7 stars \citep{Sota2011}, involving the ratio of He\,{\sc i} $\lambda$4388 to He\,{\sc ii} $\lambda$4541 or He\,{\sc i} $\lambda$4144 to He\,{\sc ii} $\lambda$4200, are used for stars with extremely weak Si diagnostics. Luminosity classes are obtained from  line morphologies of H$\gamma$ and H$\delta$ with respect to reference stars. Unusual B-type systems with Balmer emission lines and forbidden [Fe\,{\sc ii}] lines are classified as supergiant B[e] (sgB[e]) following \citet{Lamers1998} and  \citet{Kraus2019}.

For A and F supergiants, we are unable to follow the Ca\,{\sc ii} H+K criteria of \citet{Evans2003}, because our dataset excludes the Fraunhofer K line, and so again we selected SMC reference A0--5 supergiants from archival VLT/FLAMES spectroscopy, with luminosity classes assigned from the equivalent width of H$\gamma$ from comparison with \citet{Millward1985}. Spectral types assigned to late A and F stars are relatively coarse, and are primarily based on the strength of the CH G-band in BLOeM datasets spectrally degraded to the resolution of SMC AF supergiants presented by \citet{Evans2003}.  All F stars have luminosity class II--I owing to the strength of their metallic features \citep{Gray2009}, even at the low metallicity of the SMC.

The derived spectral types are compiled in Table\,\ref{tab:SpT} for each of our targets. While a multiplicity analysis is beyond the scope of this paper, a few dozen ($\approx 40/929$) already portray clear evidence for the presence of at least two stellar components in their spectra. 
These are visually identified and, when possible, a preliminary spectral type for the companion(s) is provided.

\section{Results}
\label{sec:results}

\subsection{Census}
\label{subsec:census}

\begin{figure}
\centering
\includegraphics[width=0.5\textwidth]{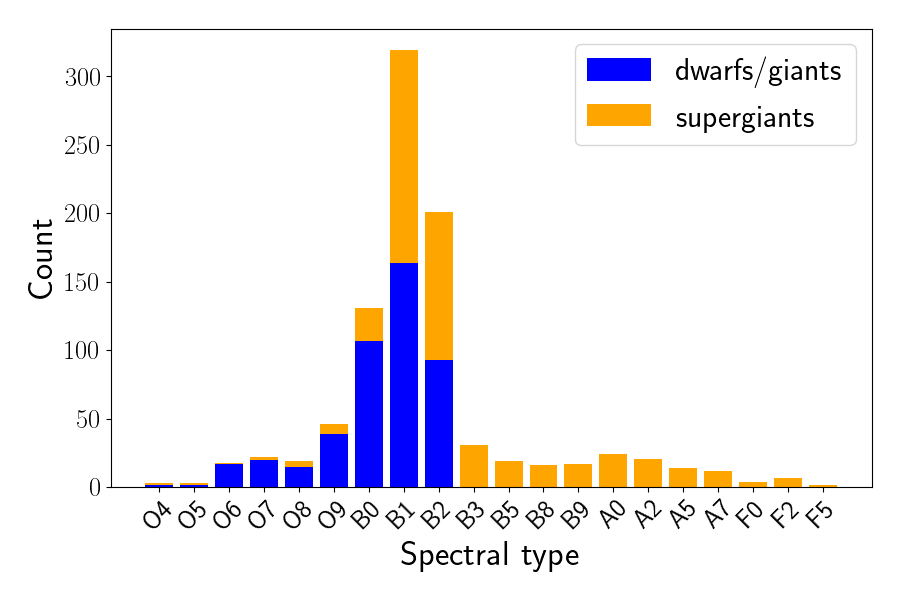}
\caption{Spectral-type histogram of the BLOeM sample. Fractional spectral types are rounded off. The sample is subdivided into dwarfs and giants (luminosity classes V-III), which are thought to be primarily core H-burning, and supergiants (luminosity classes I or II) thought to be mainly post main sequence.  }
\label{fig:SpTHisto}
\end{figure}

A histogram of the spectral-type distribution across the range O4 to F5, subdivided into dwarfs and giants in one group and supergiants in the other, is shown in Fig.\,\ref{fig:SpTHisto}. The BLOeM sample  includes the complete spectral range from the earliest stars in the SMC (excluding Wolf-Rayet stars) to yellow supergiants. Overall, the sample includes \Onum~O-type stars, \BVnum~ early B-type dwarfs and giants (B0 -- B3 V--III),  \BInum~early B-type supergiants (B0 -- B3 II--I), and \Latenum~late-type supergiants (B4 and later). Of these, \OeBenum~objects are classified as OBe stars, portraying characteristic emission in their Balmer lines: \Oenum~Oe stars (including four uncertain cases, marked `e?'), and \Benum~are Be stars (Sect.\,\ref{subsec:GaiaLowRes}).

The number of dwarfs and giants (generally interpreted as main sequence objects) steeply decreases around a spectral type of B2, which roughly corresponds to $8\,M_\odot$ \citep{Harmanec1988}. Among the B1-B2 dwarfs and giants, the vast majority are giants, bright enough to have been included in our survey. These are likely evolved main sequence stars. In the O-star regime, we are sensitive to the full extent of the main sequence down to the youngest stars, assuming those are not dust-enshrouded.

\subsection{OBe fraction: Complementary {\it Gaia} low-resolution spectra}\label{subsec:GaiaLowRes}

\begin{figure}
\centering
\includegraphics[width=0.54\textwidth]{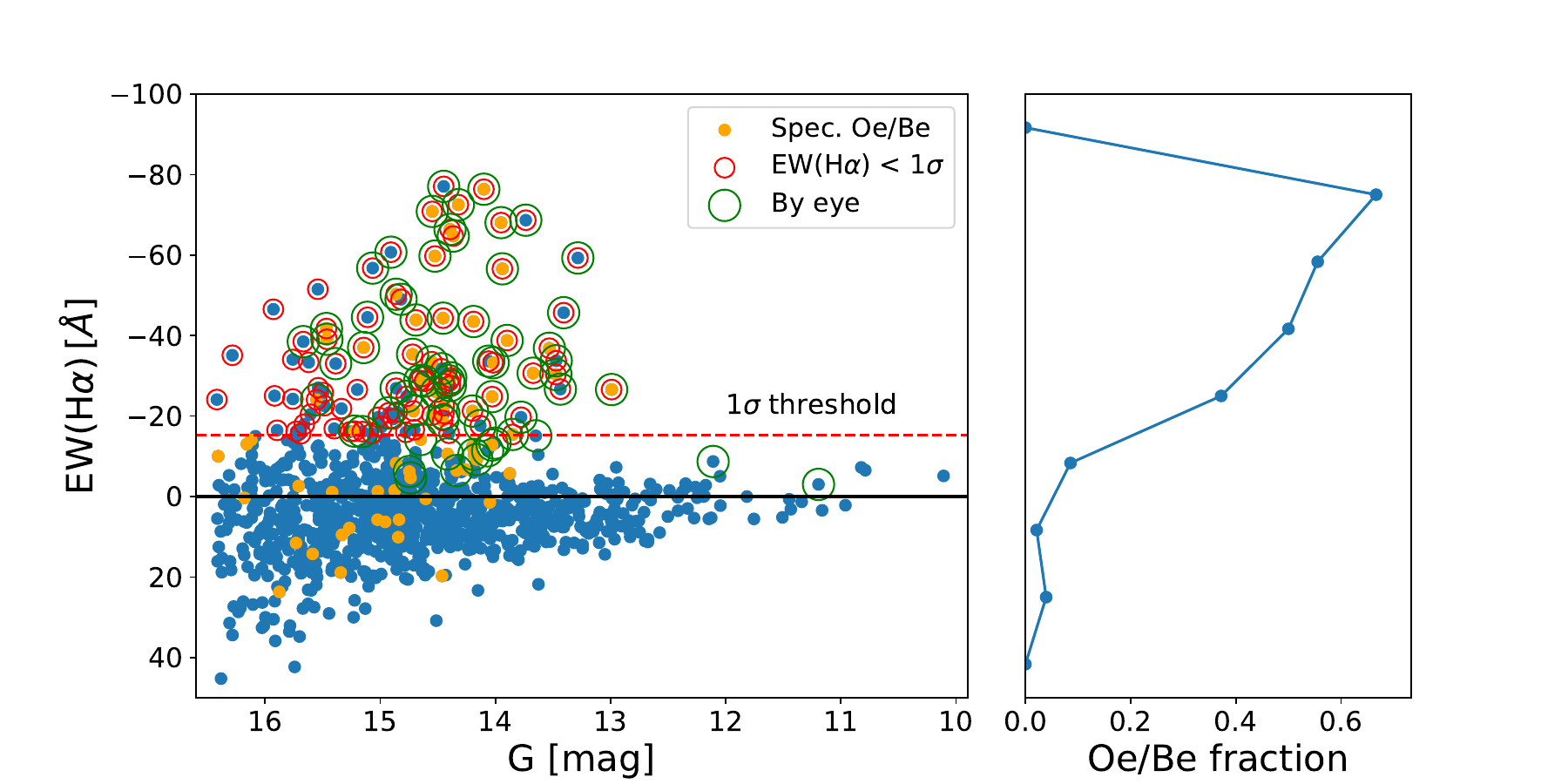}
\caption{Selection of OBe stars with {\it Gaia} data. H$\alpha$ equivalent widths of the BLOeM targets measured from {\it Gaia} low-resolution spectra. Spectroscopically classified OBe stars are marked in orange. All targets with emission exceeding 1$\sigma$ are circled in red, and by-eye identifications of H$\alpha$ emitters are marked with green circles. {\it Right:} Fraction of spectroscopically classified OBe stars using the BLOeM dataset, versus H$\alpha$ EW.  }
\label{fig:GaiaHa}
\end{figure}

As noted in Sect.\,\ref{subsec:census}, \OeBenum~targets have been identified as OBe stars in our sample. However, the BLOeM dataset does not cover the H$\alpha$ line, which is the most sensitive line for circumstellar material in the visual range. For this reason, it is likely that we have missed OBe stars in the sample. To mark additional OBe candidates,  we make use of {\it Gaia} XP low-resolution flux-calibrated spectra \citep{deangeli2023}, which exist for all sources except \object{BLOeM 4-041}, \object{6-044}, \object{7-012,} and \object{8-033}. While these data are of very low resolution ($R\approx 50$), and some are affected by problematic wiggles in the data, they may be used to detect the presence of strong emission, or candidate OBe stars in our case. To be pragmatic, we adopt a straightforward approach of measuring the H$\alpha$ equivalent width (EW) in the XP `sampled' data, using broad windows on both sides of the line to define a local continuum. While more complex approaches are possible \citep[e.g.][]{weiler2023}, our method should be adequate for detecting strong H$\alpha$ emission. Candidate OBe stars are then flagged as 1$\sigma$ outliers in the distribution of EW as a function of magnitude. As verification of this approach, a visual inspection of all spectra was also performed to define a second category of `by-eye' OBe candidates. Both samples are illustrated in Fig.\,\ref{fig:GaiaHa}. The two methods agree for most targets, although given the contamination induced by the spectral wiggles on the EWs, we favour the by-eye classification of H$\alpha$ emitters.

Overall, in addition to the \OeBenum OBe stars classified using the BLOeM dataset, we identify 16 additional H$\alpha$ emitters. However,  half of these are supergiants according to our classification from the BLOeM dataset, which, by definition, are not OBe stars. Only  eight targets are potential Be stars that have not been classified as such: BLOeM 4-046, 4-096, 5-096, 5-107, 7-008, 7-094, 7-099, and 8-078. Adding those, the total  number of OBe stars in our sample is 90, amounting to a fraction among the non-supergiant OB stars of 11\%.  This fraction is much lower than the fraction of $\approx$30\% that has been found in previous work \citep{Bonanos2010, Schootemeijer2022}. The difference can be explained by the diagonal CMD cut that we use to select our sample, which favours blue stars and hence disfavours OBe stars, which are about 0.3 magnitudes redder than OB stars (\citealt{Schootemeijer2022}, their figure A.5). 
The bias mentioned above likely does not impact the Oe fraction derived here, which is 20/159 = 13\%, comparable to the 10\% fraction reported by \citet{Bonanos2010}.

\subsection{Notable targets}
\label{subsec:peculiar}

We highlight a few unique objects in our sample. \object{BLOeM 2-116}, \object{BLOeM 3-012}, and \object{BLOeM 4-055} are classified as sgB[e] stars. \object{BLOeM 2-116} (alias \object{LHA 115-S 18}) has been the subject of several studies \citep[e.g.][]{Shore1987, Clark2013, Maravelias2014}, as has \object{BLOeM 3-012} (alias \object{RMC 4}; \citealt{Zickgraf1996, Graus2012, Pasquali2000, Wu2020}). Such objects are thought to represent a brief evolutionary phase of massive stars. Their spectra resemble those of luminous blue variables (LBVs), and their origin is debated in the literature \citep[e.g.][]{Podsiadlowski2006, Clark2013}. The spectra of all these objects indicate variability. Whether or not this variability stems from binary motion is not yet clear, though \object{BLOeM 3-012} has previously been reported to be a binary  \citep{Zickgraf1996} and seems well explained as the product of a binary merger in a triple system \citep{Pasquali2000, Podsiadlowski2006, Wu2020}.

\object{BLOeM 3-031} and \object{BLOeM 5-071} show spectra that resemble those of Be + bloated stripped-star binaries such as 
\mbox{\object{LB-1}} \citep{Irrgang2020, Abdul-Masih2020, Shenar2020LB1, El-Badry2021} and   \object{HR~6819} \citep{Bodensteiner2020Be, Frost2022}. Such objects feature strong Balmer emission characteristic of Be stars, in combination with narrow and RV variable absorption features that stem from a putative bloated stripped-star companion. 

\object{BLOeM 2-104} and \object{BLOeM 4-039} were classified as Of?p stars by \citet{Evans2004_2dF}, which refers to the presence of emission lines associated with  N\,{\sc iii} and C\,{\sc iii} in the range 4630--4660\AA\ \citep{Walborn1972}.  We cannot independently verify this, because the BLOeM data lack this range, though we do confirm the presence of emission in the Balmer lines characteristic of such stars, and so we adopt this classification. All known Galactic Of?p stars possess strong global magnetic fields \citep{Grunhut2017}. It is therefore plausible that \object{BLOeM 2-104} and \object{BLOeM 4-039} are magnetic as well, although spectropolarimetric data are needed to verify this.

Nine targets are identified as significant X-ray emitters from a cross-match with various X-ray catalogues described in Appendix\,\ref{sec:Xmatch}, one of which may be a spurious X-ray detection (\object{BLOeM 6-116}).  Four of those have been classified as high-mass X-ray binaries (HMXBs) in the past: \object{BLOeM 2-055},  \object{2-82}, \object{BLOeM 2-116}, \object{BLOeM 4-026} and \object{BLOeM 4-113}. \object{BLOeM 3-042} may be a colliding-wind binary (CWB) given its classification (O6~III + O7.5). The origin of X-rays in \object{BLOeM 8-029} (B1\,IV:) and  \object{BLOeM 1-102} (O6~III(n)) is not yet clear. Further details regarding the cross-match process and X-ray luminosities of the objects can be found in Appendix\,\ref{sec:Xmatch}. 

Finally, cross-matching with the OGLE catalogue of photometrically variable stars in the SMC \citep{Pawlak2016}, 74 (i.e. 8\%) of the 929 BLOeM targets are found to be eclipsing binaries, while 8 are classified as ellipsoidal variables (Sect.\,\ref{subsec:OGLE}).


\begin{figure*}
\centering
\includegraphics[width=\textwidth]{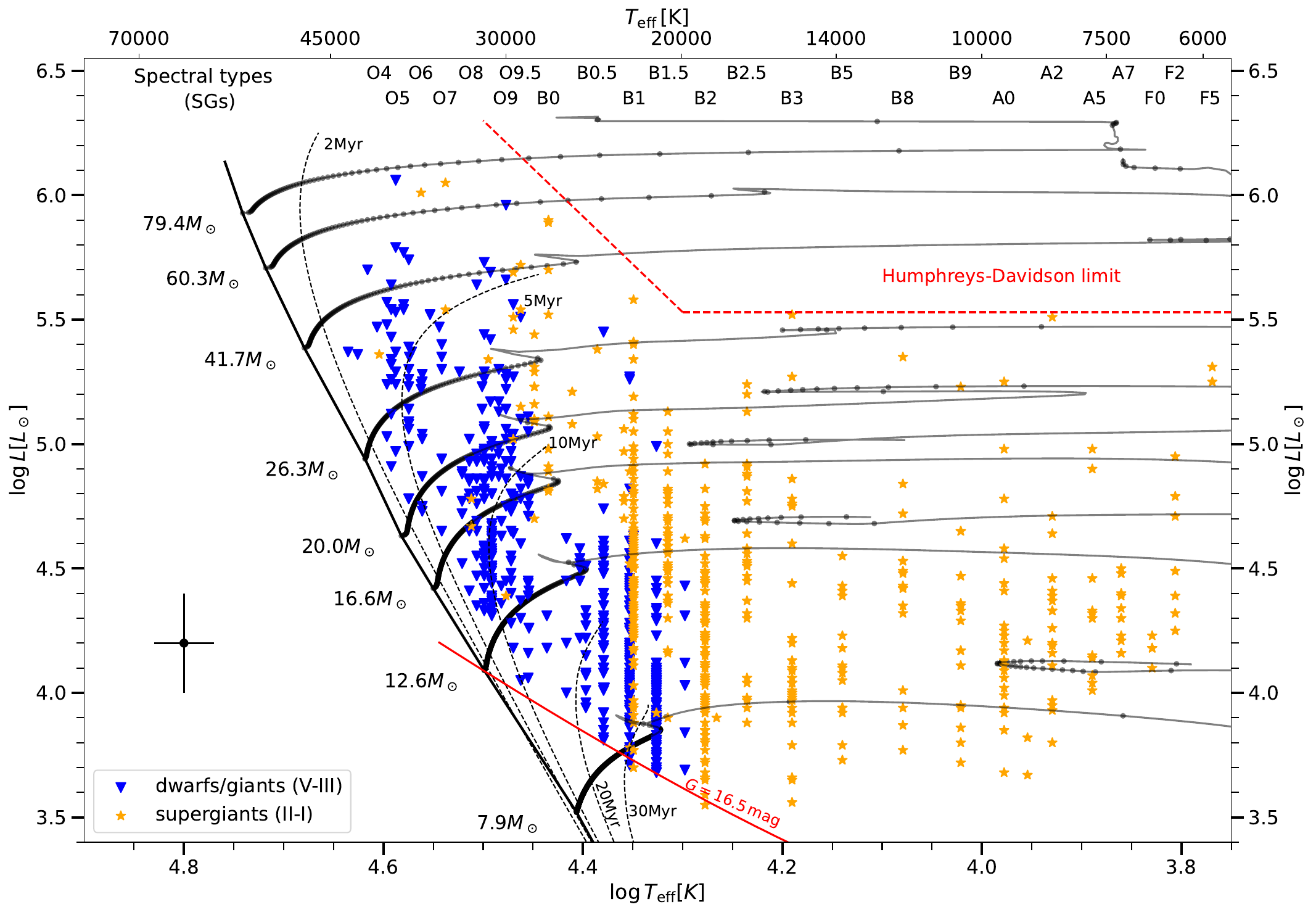}
\caption{Approximate location of the BLOeM stars in the HRD of the BLOeM sample based on
spectral-type calibrations described in Sect.~\ref{subsec:HRD}, colour-coded for dwarfs and giants (V-III) and supergiants (II-I) with blue triangles and orange stars, respectively. An estimate of the typical error (subject to calibration error, uncertain reddening, and potential binary contamination)  is shown in the bottom left corner. We use the same tracks plotted in Fig.\,\ref{fig:BLOeMOverview}, but include more initial masses (shown in labels). The $M_{\rm ini} = 7.9\,M_\odot$ track was used to select the BLOeM sample from the {\it Gaia} catalogue.  Black dots along the track are spaced by 0.05\,Myr. Also plotted are the ZAMS and isochrones at 2, 5, 10, 20, and 30\,Myr. The labels at the top (split into two rows for clarity) show the spectral types corresponding to the temperature scale for supergiants only (adopted from Table\,\ref{tab:calib}). The magnitude cut at $G=16.5\,$mag and the H--D limit (identified visually) are also marked. }
\label{fig:HRD}
\end{figure*}

\subsection{Hertzsprung--Russell diagram}
\label{subsec:HRD}

\begin{table}
\caption{Adopted spectral type--temperature calibration}\label{tab:calib}
    \begin{tabular}{l @{\hspace{5mm}} rr @{\hspace{8mm}} rr @{\hspace{8mm}} rr}
    \hline
    Spect. & \multicolumn{2}{c}{Dwarf} & \multicolumn{2}{c}{Giant} & \multicolumn{2}{c}{Supergiant}\\
   Type & $T_{\rm eff}$ & $BC_{K}$ & $T_{\rm eff}$ & $BC_{K}$ & $T_{\rm eff}$ & $BC_{K}$ \\
        & kK  & mag & kK & mag & kK & mag \\
    \hline\hline
    O4   & 46.0 & --4.99 & 41.7 & --4.87 & 40.2 & --4.78 \\
    O5   & 41.3 & --4.84 & 39.9 & --4.63 & 38.5 & --4.44 \\
    O6   & 39.5 & --4.72 & 38.0 & --4.50 & 36.5 & --4.27 \\
    O7   & 38.7 & --4.58 & 36.5 & --4.37 & 34.5 & --4.15 \\
    O8   & 36.4 & --4.30 & 35.0 & --4.15 & 32.5 & --4.01 \\
    O9   & 33.2 & --3.97 & 31.7 & --3.94 & 30.0 & --3.63 \\
    O9.5 & 32.1 & --3.82 & 30.5 & --3.75 & 29.0 & --3.48 \\
    B0   & 31.0 & --3.74 & 29.5 & --3.61 & 27.2 & --3.29 \\
    B0.5 & 29.6 & --3.57 & 28.5 & --3.47 & 24.3 & --2.89 \\
    B1   & 27.3 & --3.28 & 23.9 & --2.93 & 22.3 & --2.62 \\
    B1.5 & 26.1 & --3.17 & 22.5 & --2.70 & 20.6 & --2.38 \\
    B2   & 24.9 & --3.07 & 21.2 & --2.53 & 18.9 & --2.18 \\
    B2.5 & 23.9 & --2.95 & 19.8 & --2.34 & 17.2 & --1.78 \\
    B3   & 21.5 & --2.59 & 18.4 & --2.05 & 15.5 & --1.45 \\
    B5&\multicolumn{2}{c}{$\cdots$} & 16.7 & --1.65 & 13.8 & --1.04 \\
    B8&\multicolumn{2}{c}{$\cdots$} & \multicolumn{2}{c}{$\cdots$} & 12.0 & --0.62 \\
    B9&\multicolumn{2}{c}{$\cdots$} & \multicolumn{2}{c}{$\cdots$} & 10.5 & --0.27 \\
    A0&\multicolumn{2}{c}{$\cdots$} & \multicolumn{2}{c}{$\cdots$} & 9.5 & +0.02 \\    
    A2&\multicolumn{2}{c}{$\cdots$} & \multicolumn{2}{c}{$\cdots$} & 8.5 & +0.30 \\
    A5&\multicolumn{2}{c}{$\cdots$} & \multicolumn{2}{c}{$\cdots$} & 7.7 & +0.35 \\
    A7&\multicolumn{2}{c}{$\cdots$} & \multicolumn{2}{c}{$\cdots$} & 7.2 & +0.49 \\
    F0&\multicolumn{2}{c}{$\cdots$} & \multicolumn{2}{c}{$\cdots$} & 6.7 & +0.62 \\
    F2&\multicolumn{2}{c}{$\cdots$} & \multicolumn{2}{c}{$\cdots$} & 6.4 & +0.72 \\
    F5&\multicolumn{2}{c}{$\cdots$} & \multicolumn{2}{c}{$\cdots$} & 5.9 & +0.85 \\
    \hline
    \end{tabular}
\tablefoot{Based on SMC results from \citet{Bouret2013} and \citet{Bouret2021} for O stars, \citet{Dufton2019} for B0--5 stars, and \citet{Evans2003} for cooler supergiants. $K_{\rm s}$-band bolometric corrections combine V-band bolometric corrections from \citet{Bouret2013}, \citet{Bouret2021}, \citet{Lanz2007}, \citet{Kudritzki2008} and \citet{Cox2000}, and V-K colours from \citet{Martins2006}, \citet{Koornneef1983} and \citet{Cox2000}.}
\end{table}

Spectral classification of the BLOeM sample permits coarse estimates of surface temperatures of normal OBAF stars presented in Table~\ref{tab:calib}. These are adapted from the SMC spectral type-temperature calibration of \citet{Dufton2019}, incorporating results for SMC O stars from \citet{Bouret2013}, \citet{Bouret2021} and SMC late B and AF supergiants from \citet{Evans2003}.

Bolometric luminosities are estimated from K$_s$ band point spread function (PSF) photometry fitting from VISTA/VMC \citep{Cioni2011}, a distance modulus of 18.95 mag for the SMC \citep{Hilditch2005, Graczyk2020}, intrinsic V-K$_{s}$ colours from \citet{Martins2006}, \citet{Cox2000}, and \citet{Koornneef1983}, plus V-band bolometric corrections drawn from \citet{Bouret2013} and \citet{Bouret2021} for O stars, \citet{Lanz2007} for B stars with effective temperatures $T_{\rm eff} \geq$ 15kK, \citet{Kudritzki2008} for late B and A supergiants (8kK $\leq T \leq$ 12kK), and \citet{Cox2000} otherwise. The use of K$_s$ band photometry avoids the need to correct for interstellar extinction, which is significantly smaller at the K$_s$ band than at the $G$ band. Given this and the typically negligible extinction towards SMC sightlines \citep[see e.g.][]{Schootemeijer2021}, we assume $A_{\rm K} = 0$ here; this may affect the luminosities of individual objects, but does not affect our general conclusions. For the intermediate luminosity classes IV and II, we use the dwarf and supergiant relations from Table\,\ref{tab:calib}, respectively. 

We present the resulting Hertzsprung--Russell diagram (HRD) of the BLOeM sample in Fig.~\ref{fig:HRD}, together with evolutionary tracks and isochrones for SMC metallicity stars from the extended grid of \citet{Schootemeijer2019}. As described in Appendix B of  \citet{Hastings2021},  these models have efficient semi-convection ($\alpha_{\rm sc} = 10$) and mass-dependent overshooting ($\alpha_{\rm ov}$ linearly increases from 0.1 at $1.66\,M_\odot$ to 0.3 at 20\,$M_\odot$, and remains constant at 0.3 for higher masses). Apart from  $\alpha_{\rm sc}$ and $\alpha_{\rm ov}$, this grid has the same physics assumptions as \citet{Brott2011}.
We adopt the dwarf temperature calibration for subgiants, and the giant temperature calibration for bright giants. The BLOeM sample includes dwarfs and subgiants in the mass range 10--70 $M_{\odot}$, plus a few giants and supergiants down to $\approx$7--8 $M_{\odot}$. The three sgB[e] objects (\object{BLOeM 2-116}, \object{BLOeM 3-012}, and \object{BLOeM 4-055}) are excluded from the HRD, as standard calibrations cannot be used on them.

The errors on $\log L$ and $T_{\rm eff}$ can only be roughly estimated. The spectral types can be  roughly estimated to within one spectral bin, corresponding to $\Delta \log T_{\rm eff} {\rm[K]} \approx 0.02-0.04$; this is larger than typical calibration errors. However, the impact of binary contamination is not possible to quantify without further analysis; we set the error in Fig.\,\ref{fig:HRD} to 0.03\,dex. Similarly, errors on bolometric correction can be estimated from neighbouring spectral-type bins. Depending on the spectral type, they typically range between $\Delta BC = 0.2 - 0.4$, corresponding roughly to $\Delta \log L/L_\odot = 0.1\,$dex. Added to this are potential sources contaminating the K$_s$ magnitudes (e.g. infrared excess), binary contamination, and reddening. To remain conservative, we adopt an uncertainty of $\Delta \log L = 0.2\,$dex in Fig.\,\ref{fig:HRD}.

The HRD reveals the span of parameters covered by our sample in terms of initial mass and age. Qualitatively, it is similar to that presented by \citet{Humphreys1984}, though the BLOeM sample goes substantially deeper in magnitude ($\Delta V \approx 2.5\,$mag).
This HRD suggests that the sample probes the range $8 \lesssim M_{\rm ini} \lesssim 80\,M_\odot$, though only a minority of targets exceed initial masses of $\gtrsim 30\,M_\odot$.  In terms of age, a visual comparison with isochrones implies that the BLOeM sample probes stars as young as $\approx 2\,$Myr, extending to ages of the order of 20\,Myr or more.  The lack of very massive stars in the SMC has been noted in the past \citep[e.g.][]{Ramachandran2019, Schootemeijer2021}, and could be related to a peak of star formation $\approx 10 - 40\,$Myr ago \citep{Antoniou2010, Rubele2015, Schootemeijer2021}.  The Humphreys--Davidson (H-D) limit \citep{Humphreys1979, Humphreys1984}, which marks the absence of bright stars in the upper-right part of the HRD, is clearly seen above $\log L/L_\odot \approx 5.5$, in agreement with recent evaluations of the  H--D limit in the SMC \citep{Ramachandran2019, Davies2018, Gilkis2021, Sabhahit2021}. In Appendix \ref{sec:FieldHRD}, we show similar HRDs for each of the eight SMC fields, which provide an overview of the stellar content in each field.

We note that these estimates are subject to uncertainty given the uncertainties on $\log L$, and could also change depending on the set of evolutionary tracks being used \citep[e.g.][]{Georgy2013, Choi2016, Marigo2017, Keszthelyi2022}.  Moreover, an unknown fraction of the sample stars will be affected either by past binary interactions or bright companions, while we only make use of single-star tracks in this first characterisation effort. A full analysis of the mass and age distribution will require treatment of multiplicity (e.g. binary identification, orbital analysis, spectral disentangling), which will be the subject of subsequent papers.

\section{Conclusions}
\label{sec:conclusions}

This work presents the rationale, target selection, and first characterisation of the sample spectroscopically monitored with FLAMES/VLT in the framework of the Binarity at LOw Metallicity (BLOeM) ESO Large Programme. BLOeM will collect 25 epochs of spectroscopy of 929 massive stars in the low-metallicity conditions of the SMC in the period from October 2023 to September 2025. The sample populates eight fields within the SMC, probing several of its environments, though limited to field stars. 

The goals of the survey are to use the time-dependent RVs of all targets to derive the observed and intrinsic binary fraction as a function of stellar mass and age, derive the orbits of all identified binaries, and, through this, establish fundamental properties such as the initial mass function, star formation history, surface abundance pattern, and orbital-parameter distributions of massive stars at low metallicity. Moreover, the survey will enable the identification of unique evolved binaries, such as dormant OB+BH binaries, and provide testbeds for single-star evolution models via dynamical-mass measurements of eclipsing binaries.

The present study is based on 9/25 epochs acquired thus far in the framework of the BLOeM survey during the first semester, with our main goal being to describe the sample selection (Sect.\,\ref{sec:sample}), develop a data-reduction pipeline (Sect.\,\ref{sec:data}), perform a spectral classification (Sect.\,\ref{sec:SpT}), and investigate the mass and age domains probed by the sample (Sect.\,\ref{sec:results}).  

The sample spans a wide range of spectral types, extending from O4 for the earliest subtypes to F5 for the latest ones, though it is dominated by B0-2 stars. In total, there are 159 O-type stars, 324 early B-type (B0--B3) dwarfs, subgiants, and giants, 309 early B-type bright giants and supergiants, and 137 late-type supergiants. From spectral-type calibrations and usage of K$_s$ magnitudes, we derived the effective temperatures and luminosities of the targets, assuming they are all single stars. The sample covers the regime $6.5\lesssim T_{\rm eff} \lesssim 45$~kK and $3.7 \lesssim \log L/L_\odot \lesssim 6.1$. From comparison to evolution tracks and isochrones extended from \citet{Schootemeijer2019}, this roughly corresponds to initial masses in the range $8 - 80\,M_\odot$  and ages mainly in the range $\approx 2 - 20\,$Myr. The sample corroborates the blue region of the Humphreys-Davidson limit.

We highlight a few peculiar objects in our sample: eight sources are confirmed as X-ray bright, four of which have been classified as HMXBs in the past, and one of which is a promising CWB candidate. Three objects are classified as sgB[e]/LBV-like stars, and two as Be + bloated stripped-star binary candidates. Two candidate magnetic stars, classified as Of?p stars previously, are also included in the sample.

We identify  \OeBenum stars as OBe stars.  As we lack the diagnostic H$\alpha$ line, this should be considered a lower limit, although usage of low-resolution {\it Gaia} spectra covering the H$\alpha$ line only yields a few more candidates, amounting to an OBe fraction of $\approx 11\%$ relative to the non-supergiant OB sample. The Oe fraction among all O stars in the sample is 13\%. However, the Be fraction is merely 11\%, which is lower than reported in the literature. It is likely that we are biased against Be stars in this sample, because these tend to be redder compared to a `standard' evolutionary track, which we used to select our sample (see Sect.\,\ref{subsec:GaiaLowRes}). However, the Oe sample is likely unbiased, and so this fraction should represent the Oe fraction in the SMC. Finally,  74 eclipsing binaries are identified, as well as 8 ellipsoidal variables (Sect.\,\ref{subsec:peculiar} and Appendix\,\ref{sec:Xmatch}).
We are currently carrying out a systematic analysis of the first batch of data for RV variability for the various subsamples of the study. 


\begin{acknowledgements}

The research leading to these results has received funding from the European Research Council (ERC) under the European Union's Horizon 2020 research and innovation programme (grant agreement numbers 772225: MULTIPLES). PAC and JMB
are supported by the Science and Technology Facilities Council
research grant ST/V000853/1 (PI. V. Dhillon).
DMB gratefully acknowledges support from UK Research and Innovation (UKRI) in the form of a Frontier Research grant under the UK government's ERC Horizon Europe funding guarantee (SYMPHONY; PI Bowman; grant number: EP/Y031059/1), and a Royal Society University Research Fellowship (PI Bowman; grant number: URF{\textbackslash}R1{\textbackslash}231631). ZK acknowledges support from JSPS Kakenhi Grant-in-Aid for Scientific Research (23K19071).  IM acknowledges support from the Australian Research Council (ARC) Centre of Excellence for Gravitational Wave Discovery (OzGrav), through project number CE230100016. 
AACS, VR, RRL, and MBP are funded by the Deutsche Forschungsgemeinschaft (DFG, German Research Foundation) in the form of an Emmy Noether Research Group -- Project-ID 445674056 (SA4064/1-1, PI Sander). GGT and JJ are supported by the German Deutsche Forschungsgemeinschaft (DFG) under Project-ID 496854903 (SA4064/2-1, PI Sander)
VR, GGT, and AACS further acknowledge support from the Federal Ministry of Education and Research (BMBF) and the Baden-W{\"u}rttemberg Ministry of Science as part of the Excellence Strategy of the German Federal and State Governments. ECS acknowledges financial support by the Federal Ministry for Economic Affairs and Climate Action (BMWK) via the German Aerospace Center (Deutsches Zentrum f\"ur Luft- und Raumfahrt, DLR) grant 50 OR 2306 (PI: Ramachandran/Sander).
This work has received funding from the European Research Council (ERC) under the European Union's Horizon 2020 research and innovation programme (Grant agreement No.\ 945806) and is supported by the Deutsche Forschungsgemeinschaft (DFG, German Research Foundation) under Germany's Excellence Strategy EXC 2181/1-390900948 (the Heidelberg STRUCTURES Excellence Cluster).  LMO is thankful for the funding provided by the DFG grant 443790621. This paper benefited from discussions at the International Space Science Institute (ISSI) in Bern through ISSI International Team project 512 (Multiwavelength View on Massive Stars in the Era of Multimessenger Astronomy). DP acknowledges financial support by the Deutsches Zentrum f\"ur Luft und Raumfahrt (DLR) grant FKZ 50OR2005. JIV acknowledges the European Research Council for support from the ERC Advanced grant ERC-2021-ADG101054731.  JSV is 
supported by STFC (Science and Technology Facilities Council) funding under grant number ST/V000233/1. 
GH, SS-D, SRB and AH acknowledge support from the State Research Agency (AEI) of the Spanish Ministry of Science and Innovation (MICIN) and the European Regional Development Fund, FEDER under grants PID2021-122397NB-C21 and CEX2019-000920-S. SRB also acknowledges financial support by  NextGeneration EU/PRTR and MIU (UNI/551/2021) through grant Margarita Salas-ULL.
DFR is thankful for the support of the CAPES-Br and FAPERJ/DSC-10 (SEI-260003/001630/2023).
F.N., and L.R.P. acknowledge support by grants
PID2019-105552RB-C41 and PID2022-137779OB-C41 funded by
MCIN/AEI/10.13039/501100011033 by "ERDF A way of making
Europe".
MG acknowledges financial support from the grants PID2021-125485NB-C22,  CEX2019-000918-M funded by MCIN/AEI/10.13039/501100011033 (State Agency for Research of the Spanish Ministry of Science and Innovation) and SGR-2021-01069 (AGAUR).
GM acknowledges funding support from the European Research Council (ERC) under the European Union's Horizon 2020 research and innovation programme (Grant agreement No. 772086).
JMA acknowledges support from the Spanish Government Ministerio de Ciencia e Innovaci\'on and Agencia Estatal de Investigaci\'on (10.13\,039/501\,100\,011\,033) through grant PID2022-136\,640~NB-C22 and from the Consejo Superior de Investigaciones Cient\'ificas (CSIC) through grant 2022-AEP~005. 
MP is supported by the BEKKER fellowship BPN/BEK/2022/1/00106 from the Polish National Agency for Academic Exchange.
KS is funded by the National Science Center (NCN), Poland, under grant number OPUS 2021/41/B/ST9/00757.
JM acknowledges support from a Royal Society--Science Foundation Ireland University Research Fellowship.
SJ acknowledges support from the FWO PhD fellowship under project 11E1721N. 
FB acknowledges the support of the European Research Council (ERC) Horizon Europe under grant agreement number 101044048.

\end{acknowledgements}

\bibliographystyle{aa}
\bibliography{papers}

\begin{appendix}

\section{Epoch dates and spectral types}
\label{sec:EpochSpT}

\begin{table*}[ht]
\small
\caption{Field centres and epoch dates for the eight fields.}
\label{tab:MJDs}
\centering
\begin{tabular}{lcccccccc}
\hline \hline
 & Field 1 & Field 2 & Field 3 & Field 4 & Field 5 & Field 6 & Field 7 & Field 8 \\
\hline
RA centre [h:m:s] &01:02:53.8 & 00:52:01.0 & 00:48:20.2 & 00:59:41.8  & 01:05:56.2 & 01:15:08.2 & 00:59:51.4 & 00:52:44.2\\
DEC centre [d:':''] &-72:05:26.1 & -72:41:26.1 & -73:16:14.1 & -72:11:26.1 & -72:19:50.1 & -73:15:02.1 & -72:37:50.1 & -72:15:02.1\\
\hline
Epoch 1 [MJD] & 60219.10 & 60242.07 & 60242.10 & 60242.12 & 60242.15 & 60261.16 & 60246.12 & 60246.14\\
Epoch 2 [MJD] & 60220.26 & 60247.19 & 60247.21 & 60247.24 & 60247.26 & 60267.20 & 60248.17 & 60261.18\\
Epoch 3 [MJD] & 60242.05 & 60256.20 & 60256.24 & 60254.35 & 60256.22 & 60280.02 & 60256.18 & 60267.06\\
Epoch 4 [MJD] & 60245.02 & 60261.09 & 60261.11 & 60256.27 & 60261.03 & 60282.12 & 60261.07 & 60268.11\\
Epoch 5 [MJD] & 60248.15 & 60267.15 & 60262.13 & 60261.14 & 60262.11 & 60285.09 & 60267.04 & 60280.12\\
Epoch 6 [MJD] & 60252.21 & 60281.10 & 60267.09 & 60267.12 & 60267.17 & 60286.22 & 60268.13 & 60280.14\\
Epoch 7 [MJD] & 60256.29 & 60285.16 & 60270.03 & 60281.05 & 60270.05 & 60288.14 & 60280.16 & 60282.10\\
Epoch 8 [MJD] & 60261.05 & 60289.05 & 60281.08 & 60285.18 & 60280.04 & 60290.05 & 60286.19 & 60285.11\\
Epoch 9 [MJD] & 60285.20 & 60290.09 & 60285.13 & 60289.07 & 60281.12 & 60291.09 & 60288.16 & 60288.19\\
\hline
\end{tabular}
\end{table*}

Table\,\ref{tab:MJDs} provides the field centres and MJD values of the mid-exposures of the combined two sub-exposures per field and epoch. Table\,\ref{tab:SpT} provides {\it Gaia} DR3 IDs, common aliases, K$_s$ magnitudes from the VISTA/VMC catalogue \citep{Cioni2011}, previous spectral-type classifications, newly derived spectral types, flags for H$\alpha$-emitters in low-resolution {\it Gaia} spectra (Sect.\,\ref{subsec:GaiaLowRes}), and additional comments. Newly derived spectral types are based on the BLOeM dataset unless otherwise stated in the comments.

\onecolumn

\begin{centering}
\fontsize{7.pt}{7pt}\selectfont 
\setlength{\tabcolsep}{2pt}
\begin{longtable}{l c l c l l c c}
\captionsetup{width=\linewidth}
\caption{Provided from left to right are BLOeM IDs, Gaia DR IDs, common aliases, K$_s$ magnitudes from the VISTA catalogue \citep{Cioni2011}, spectral types from the literature (references are provided at the end of the table), newly derived spectral types, H$\alpha$ emitters based on low-resolution {\it Gaia} spectroscopy (see Sect.\,\ref{subsec:GaiaLowRes}), and additional comments.}\\
\hline
ID  & Gaia DR3 ID & Alias &   K$_s\,$[mag] & SpT (old)   & SpT (new)   & H$\alpha$ emit.\ & Comment \\
\hline
1-001 & 4690503998385774848 & AzV 76F & 13.69 & B8 [Azz79] & B9 Iab & no &  - \\
1-002 & 4690519082313236608 & [M2002] SMC 49825 & 15.85 & B0.2  III [Lam16] & B0 IV: & no &  - \\
1-003 & 4690519151032663296 & - & 15.52 & - & B2.5 II: & no &  - \\
1-004 & 4690501146531455744 & Sk 92, AzV 260 & 13.57 & B1.5  Iab [Len97] & B1 Ib & no &  - \\
1-005 & 4690505510214025472 & [M2002] SMC 50031 & 15.45 & B0 [She13] & B1 II & no &  - \\
1-006 & 4690520010026062336 & OGLE SMC725.15.018759 & 15.12 & - & O9.5 V + early B & no & SB2, EB [Paw13]\\
\end{longtable}
 
Ard77: \citet{Ardeberg1977}; 
Azz79: \citet{Azzopardi1979}; 
Azz82: \citet{Azzopardi1982}; 
Coe15: \citet{Coe2015}; 
Eva04: \citet{Evans2004_2dF}; 
Eva06: \citet{Evans2006}; 
Gar87: \citet{Garmany1987}; 
Gra12: \citet{Graus2012}; 
Hil05: \citet{Hilditch2005}; 
Hum83: \citet{Humphreys1983}; 
Hum91: \citet{Humphreys1991}; 
Lam13: \citet{Lamb2013}; 
Mar07: \citet{Martayan2007}; 
Mas95: \citet{Massey1995}; 
Mas00: \citet{Massey2000}; 
Mas02: \citet{Massey2002}; 
Mas04: \citet{Massey2004}; 
Mas09: \citet{Massey2009}; 
Men06: \citet{Mennickent2006}; 
Mok06: \citet{Mokiem2006}; 
Mor03: \citet{Morrell2003}; 
Neu10: \citet{Neugent2010}; 
Neu18: \citet{Neugent2018}; 
Pau12: \citet{Paul2012}; 
Paw16: \citet{Pawlak2016}; 
Pri87: \citet{Prinja1987}; 
San68: \citet{Sanduleak1968}; 
She13: \citet{Sheets2013}; 
Smi97: \citet{SmithNeubig1999}; 
Tes87: \citet{Testor1987}; 
Tes01: \citet{Testor2001}; 
Wal83: \citet{Walborn1983}; 
Wal00: \citet{Walborn2000}; 
Wal02: \citet{Walborn2002_FUSE}; 
Zas14: \citet{Zasche2014}; 
Zic96: \citet{Zickgraf1996}; 

\end{centering}

\twocolumn

\section{DSS images for nebular contamination}\label{sec:DSS}
In order to get a better handle on which stars are affected by nebular contamination, and which objects show intrinsic emission like in the case of classical OeBe stars, we investigated wide-field Digitized Sky Survey (DSS)\,2-red images. We retrieved 25'\,x\,25' cutouts for each of the fields and overplotted all \bloem\ sources in order to investigate their local surroundings. We inspected the spectra of all sources that show overdensitites in DSS\,2-red, in particular all objects classified as emission-line stars, to better distinguish between nebular contamination and classical OeBe stars. We designated with 'neb' all  objects located inside a nebulosity visible in DSS\,2-red and show narrow emission lines, mainly in the H$\gamma$ line.

\begin{figure*}
\centering
\begin{tabular}{cc}
\includegraphics[width=0.5\textwidth]{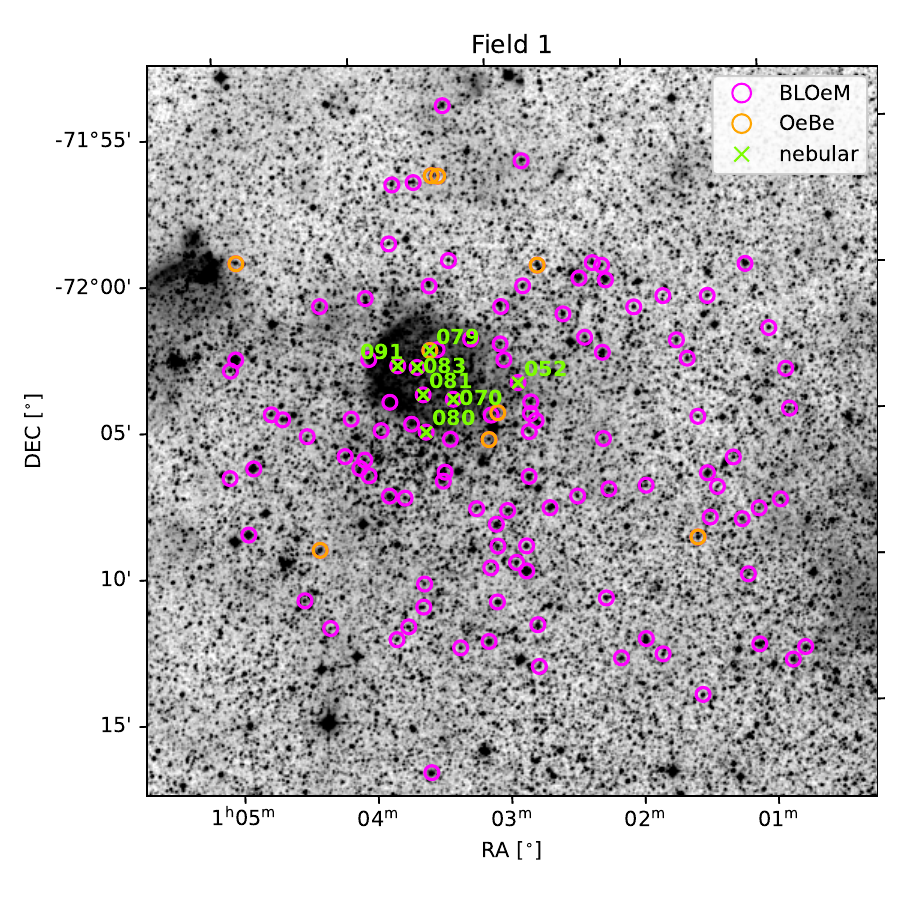} & 
\includegraphics[width=0.5\textwidth]{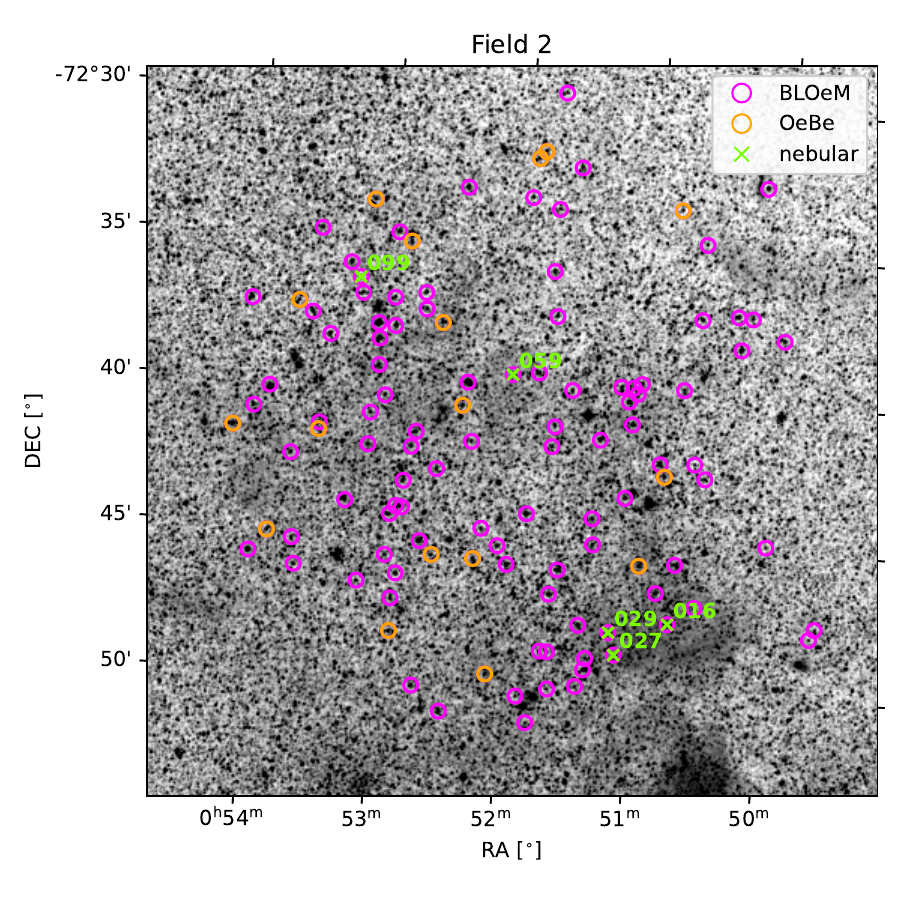} \\ 
\includegraphics[width=0.5\textwidth]{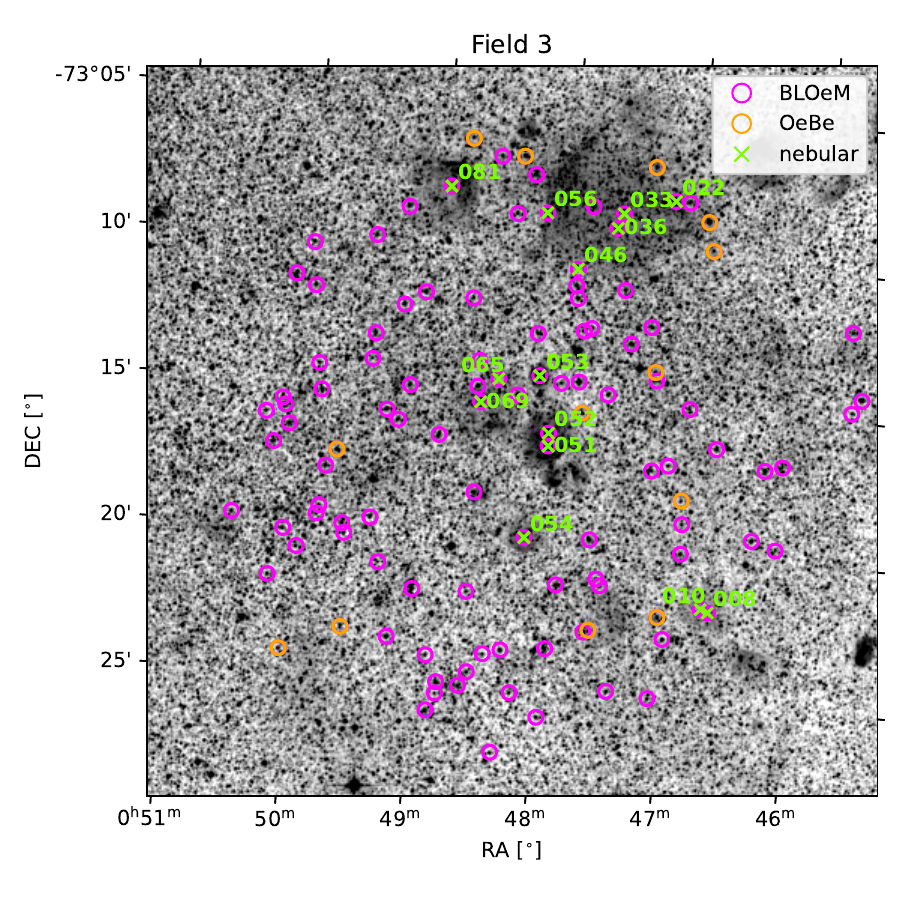} & 
\includegraphics[width=0.5\textwidth]{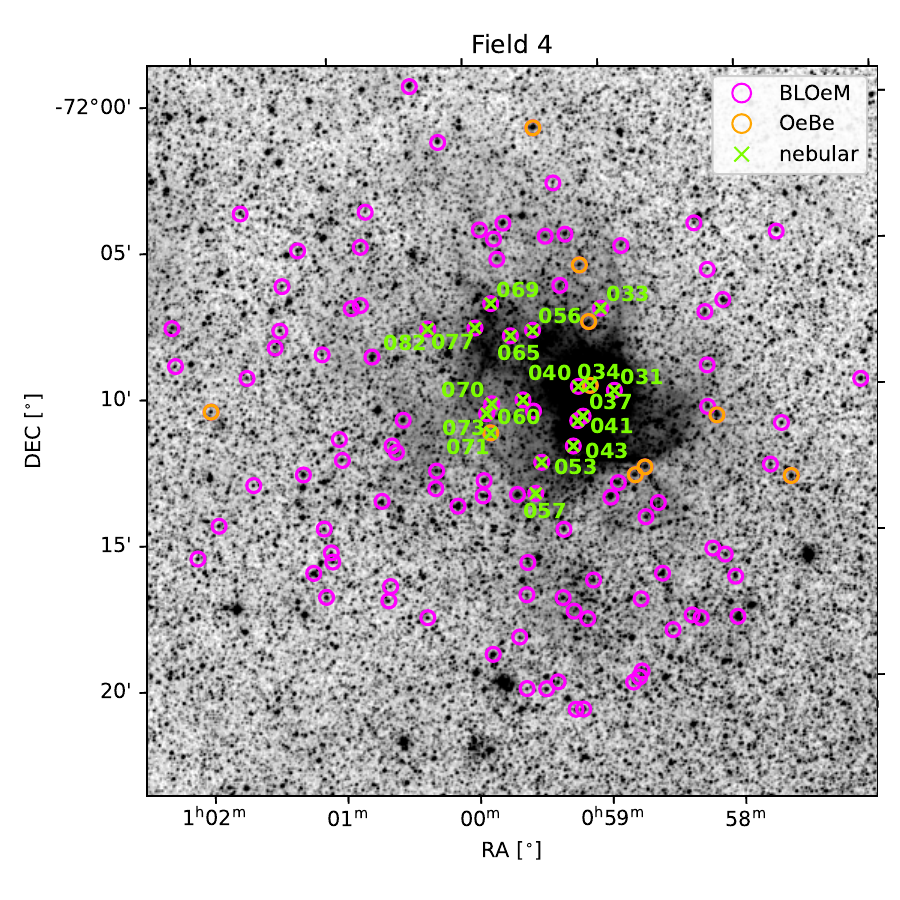} \\ 
\end{tabular}
\caption{\bloem\ sources in Field 1--4 overlaid on a DSS\,2-red image (pink circles). OeBe stars are marked with orange circles. Stars that show nebular contamination in their spectra are marked with green crosses.}
\label{fig:Fields1-4}
\end{figure*}

\begin{figure*}
\centering
\begin{tabular}{cc}
\includegraphics[width=0.48\textwidth]{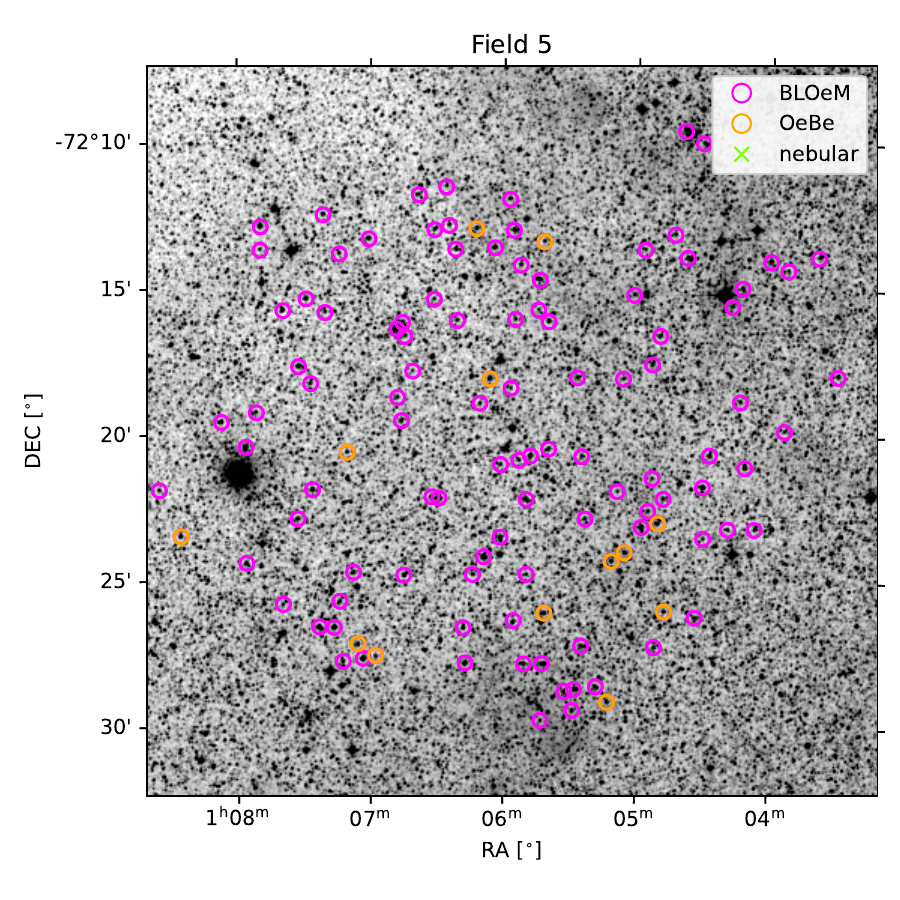} & 
\includegraphics[width=0.48\textwidth]{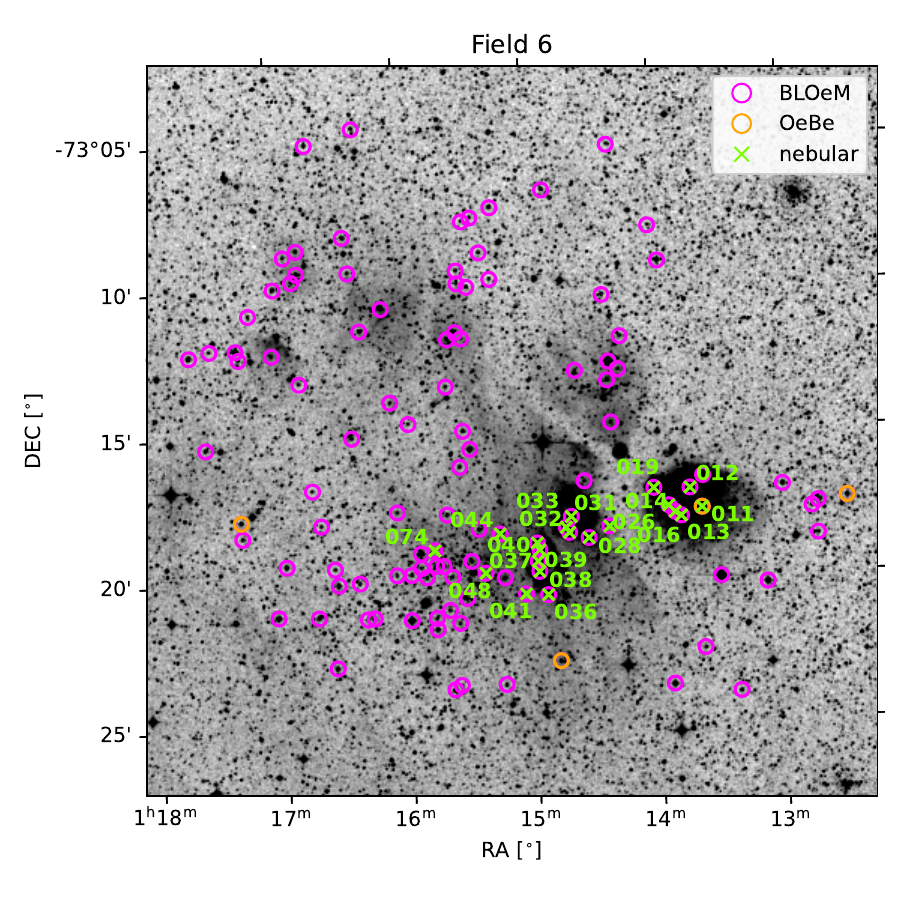} \\ 
\includegraphics[width=0.48\textwidth]{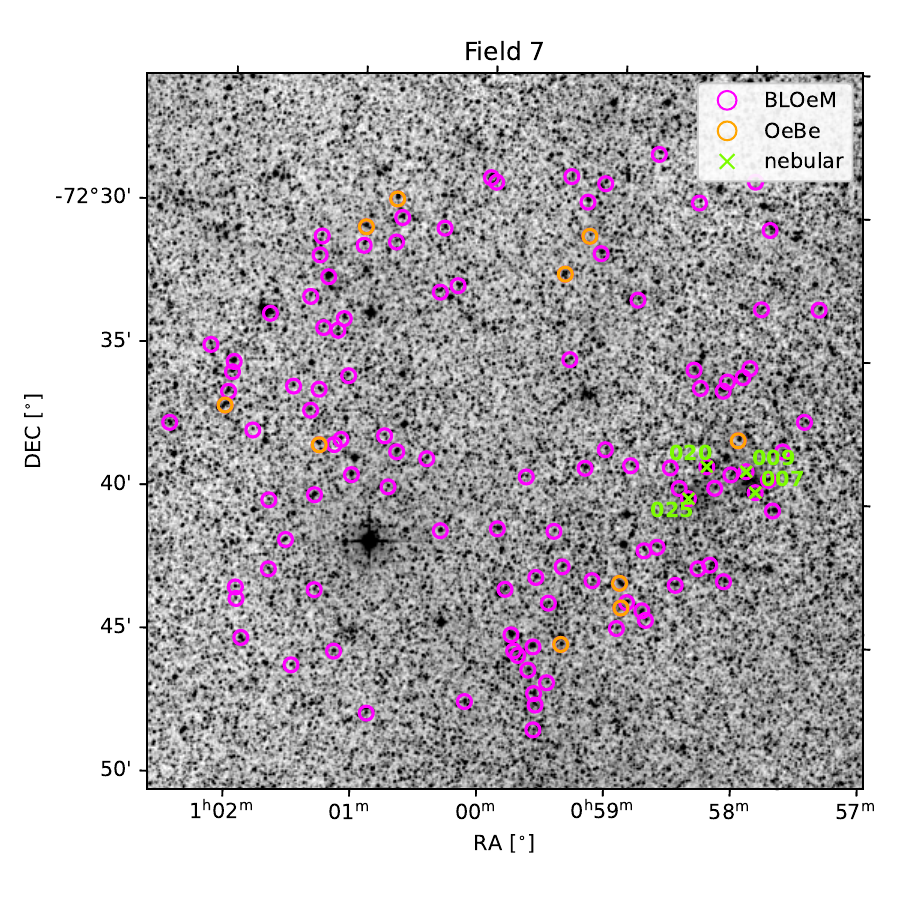} & 
\includegraphics[width=0.48\textwidth]{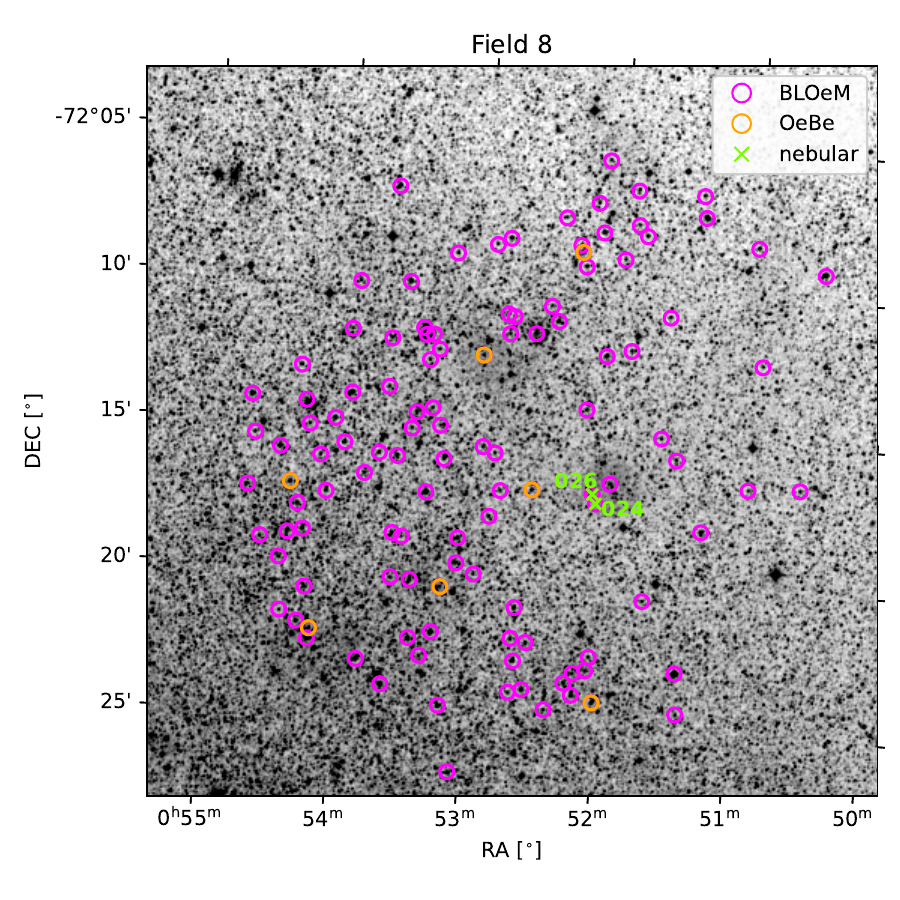} \\ 
\end{tabular}
\caption{Same as Fig.\,\ref{fig:Fields1-4}, but for fields 5 -- 8. } 
\label{fig:Fields5-8}
\end{figure*}

In Figs. \ref{fig:Fields1-4} and \ref{fig:Fields5-8} we show the DSS\,2-red cutouts for each of the eight fields. Here, we overplot all \bloem\ sources, in particular emission-line stars classified as OeBe stars, and mark stars that are affected by nebular contamination. Some fields, for example Field 1, 4 or 6, have large nebulosities in the fields of view and many sources are affected by nebular contamination. Other fields, like fields 5, 7 or 8, are barely or not affected at all. Few objects are classical OeBe stars and additionally show nebular contamination in their spectra.

\section{Cross-matches with additional catalogues}
\label{sec:Xmatch}

\subsection{ESO archive}
\label{subsec:ESO}

We cross-matched the BLOeM catalogue with spectroscopic databases in the ESO archive, defining a search radius of 3'' per target. We retrieve a total of 1,988 spectra for 202 stars out of the 929 in our sample. These spectra were acquired with various instruments of the VLT. 
These data will be used to improve the quantitative spectroscopy in subsequent papers. 

\subsection{Hubble UV Legacy Library of
Young Stars as Essential Standards (ULLYSES)}
\label{subsec:ULLY}

ULLYSES is a legacy survey of the \textit{Hubble} Space Telescope (HST), which includes the acquisition of high-resolution UV spectra for 128 massive stars in the SMC\footnote{https://ullyses.stsci.edu/ullyses-targets-smc.html}. The programme also includes a follow-up with the X-SHOOTER spectrograph of the VLT to obtain a visual and infrared coverage of the targets \cite{Vink2023}. A cross-match of the BLOeM sample with the ULLYSES sample, using a search radius of 3'', resulted in an overlap of 43 targets. From ULLYSES and XShootU there will be broad wavelength coverage of the UV and visible spectrum of these objects that will also be used to inform the analysis of these targets.

\subsection{X-ray catalogues}

The SMC was extensively observed in X-rays. The largest modern X-ray observatories,
{\em XMM-Newton} and {\em Chandra}, which operate in 0.2-12.0\,keV range,
conducted surveys of the entire SMC galaxy \citep{Laycock2010,Sturm2013}.
The deep observations of individual fields, such as the SMC Wing and NGC\,346
star cluster have also been performed \citep{Naze2004,Oskinova2013}. However,
despite these efforts X-ray emission of individual `normal' massive OB stars
is below current detection limits. On the other hand, X-ray detections of massive stars
in the SMC allow to select binary stars. Specifically, X-ray detections
are excellent tracers of CWBs,
some of which are significantly more X-ray bright compared to single stars
\citep{Corcoran1996, Sana2006, Oskinova2005,Naze2007}. But best of all, X-ray detections are suited to
identify high-mass X-ray binaries (HMXBs), where a compact object is accreting matter
of its OB-type companion. HMXBs are X-ray variable, especially so are BeXRBs where the
donor stars have OBe spectral type. BeXRBs are transient X-ray sources, and may remain
quiescent over long periods of time and could be detected only during outbursts.

To explore X-ray properties of our targets, the BLOeM catalog was cross-correlated
with catalogues produced by the {\em XMM-Newton}, {\em Chandra}, eROSITA, and  ROSAT X-ray telescopes. Only eight BLOeM
stars are firmly detected, while the positional uncertainty of one X-ray source
(6-116) precludes its firm detection.  The detected sources are listed in
Table\,\ref{tab:xray}. There are four already known HMXBs among them. Four other
X-ray sources may be either CWBs or newly discovered HMXBs. The BLOeM spectroscopy
will shed light on their nature, since HMXBs are SB1 systems while CWBs are likely
to be SB2. Some sources listed in Table\,\ref{tab:xray} have different fluxes according
to different catalogues. This may reflect true source variability, e.g.\ in case of
BeXRBs. To estimate the X-ray luminosity, we adopted a neutral hydrogen
column density $N_{\rm H}=5\times 10^{21}$\,cm$^{-2}$  and a power-law
spectrum with $\Gamma=1.7$ for all objects.

\begin{centering}
\setlength{\tabcolsep}{2pt}

\begin{table*}
\caption[ ]{BLOeM stars detected in X-rays.}
\label{tab:spec}
\footnotesize
\begin{tabular}[]{ccccccccc}
\hline
\hline
  BLOeM ID  & Alias & Uncertainty & Separation  &  $F_{\rm X}$  &  Spectral Type  &  $L_{\rm X}$ &  Catalog & Remarks\\
& &  X-ray ($''$) &   ($''$) & [erg\,cm$^{-1}$\,s$^{-1}$] &   & (erg\,s$^{-1}$) &  &  \\
\hline
3-042  & AzV 26 & 1.9  & 0.7 & $5.2 \pm 1.9 \times 10^{-16} $ & O6~I(f)+O7.5 & 4e32 &  CSC v.2   & CWB (?) \\ \hline
2-055  & AzV 102 & 0.95 & 0.7 & $30 \pm 3 \times 10^{-15}$  &  O9.7 V:n e    & 2e34 & SMCDFSCXO  & HMXB SXP 8.80 \\
       &  & 1.0  & 1.8 & $3.6\pm 2.6 \times 10^{-15}$ &            & 3e33 & 4XMM-DR13s & \\
       & & 1.3  & 0.6 & $1.8 \times 10^{-13}$           &            & 1e35 & CXOGSGSRC  & \\
       & & 2.6  & 0.2 & $29 \pm 4 \times 10^{-15}$ &            & 2e34 & CSC v.2    & \\ \hline
8-029  & - &  1.9  & 0.6  & $4.3 \pm 1.5 \times 10^{-14}$ & B1 IV:     & 3e34 & 4XMM-DR13  & HMXB (?) \\ \hline
2-082  & AzV 138 &  1.7  & 0.4  & $87 \pm 5 \times 10^{-16}$ & O9.2 III pe  & 6e33 &  CSC v.2 & HMXB \\ \hline
2-116  & AzV 154 &  0.9  & 2.4  & $5.1 \pm 2.9 \times 10^{-15}$ & sgB[e]     & 4e33 &  4XMM-DR13s & HMXB \\
       & & 1.4  & 1.2  & $8.4 \pm 4.5 \times 10^{-15}$  &           & 6e33 & SMCPSCXMM &  \\
       & &      & 0.6  & $14 \pm 6 \times 10^{-16}$  &           & 1e33 &  CSC v.2 & \\ \hline
4-026  &  Cl* NGC 346 MPG 217 &    & 0.1  & $7.1 \pm 1.5 \times 10^{-16}$  &  O9.5 IIIpe   & 5e32 & CSC v.2 & HMXB  \\ \hline
4-113  & OGLE SMC-SC9 131970 & 0.86 & 1.2 & $25 \pm 9 \times 10^{-15}$ &   B2.5 II pe    & 2e34 &   SMCPSCXMM & HMXB \\
       & &1.5  & 1.2 & $3.6 \times 10^{-14}$          &             & 3e34 &  CXOGSGSRC &  \\
       & &     & 1.7 &                  &             &      & CSC v.2 & \\ \hline
1-102  & AzV 345a &      & 0.1 & $2.07 \times 10^{-16}$          & O6 III(n)        & 1e32 &  CSC v.2 & CWB ? HMXB?\\ \hline
6-116  & 2dFS 3274 &  1.7 & 1.4 & $9.6 \pm 6.0 \times 10^{-15}$  & A7 Iab     & 7e33 & SMCPSCXMM & spurious? \\
       & & 1.1 & 2.5 & $15 \pm 7 \times 10^{-15}$  &            & 1e34 & 4XMM-DR13s & \\
\hline
\end{tabular}
\tablefoot{The columns are, in order of appearance: BLOeM identifiers, uncertainties on the X-ray position from the corresponding X-ray catalog, separations between {\it Gaia} DR3 coordinates and X-ray coordinates, 
X-ray fluxes, energy ranges from corresponding catalogs, 
spectral type, 
the seventh column: estimated X-ray luminosity in the same energy range as flux;
the eighth column: catalog name;
the ninth column: preliminary identification of a source type. The catalogues are:}
\begin{list}{}{}
\item[CSC v.2] \textit{Chandra} Source Catalog v.2 \citep{cscv2}
\item[SMCDFSCXO] SMC Deep Fields X-Ray Point Source Catalog \citep{Laycock2010}
\item[4XMM-DR13s] \textit{XMM-Newton} Serendipitous Source Catalog from Stacked Observations\citep{Traulsen2020}
\item[4XMM-DR13]\textit{XMM-Newton} Serendipitous Source Catalog DR13 \citep{ Webb2020}
\item[CXOGSGSRC] \textit{Chandra} ACIS GSG Point-Like X-Ray Source Catalog \citep{Wang2016}
\item[SMCPSCXMM] SMC \textit{XMM-Newton} Point Source Catalog \citep{Sturm2013}
\end{list}
\label{tab:xray}
\end{table*}

\end{centering}

\subsection{OGLE photometry}
\label{subsec:OGLE}

A large number of the BLOeM targets has been monitored by the OGLE photometric survey. Out of the 929 targets, 847 were observed in the OGLE-III \citep{Udalski2003} and 785 continue to be observed in the OGLE-IV \citep{Udalski2015}. Among them, there are 82 objects identified as binary systems by \citet{Pawlak2016}, including 74 eclipsing and 8 ellipsoidal binaries.

\subsection{TESS photometry}
\label{subsec:TESS}

All BLOeM targets have been and continue to be periodically observed by the all-sky time-series photometry TESS space mission \citep{Ricker2015}. An initial survey of photometric variability for XShootU targets in the LMC and SMC was performed by \citet{Bowman2024}, who found similar stochastic low-frequency (SLF) variability to Galactic OB stars \citep{Bowman2019}. In the future, we will extract light curves for as many BLOeM targets as possible to identify possible eclipsing binaries, stars with rotational modulation, and pulsations. In so doing, this will provide complementary constraints and inform the search for multiplicity for the BLOeM sample.

\section{HRDs separated by field}
\label{sec:FieldHRD}

\begin{figure*}
\centering
\begin{tabular}{cc}
\includegraphics[width=0.45\textwidth]{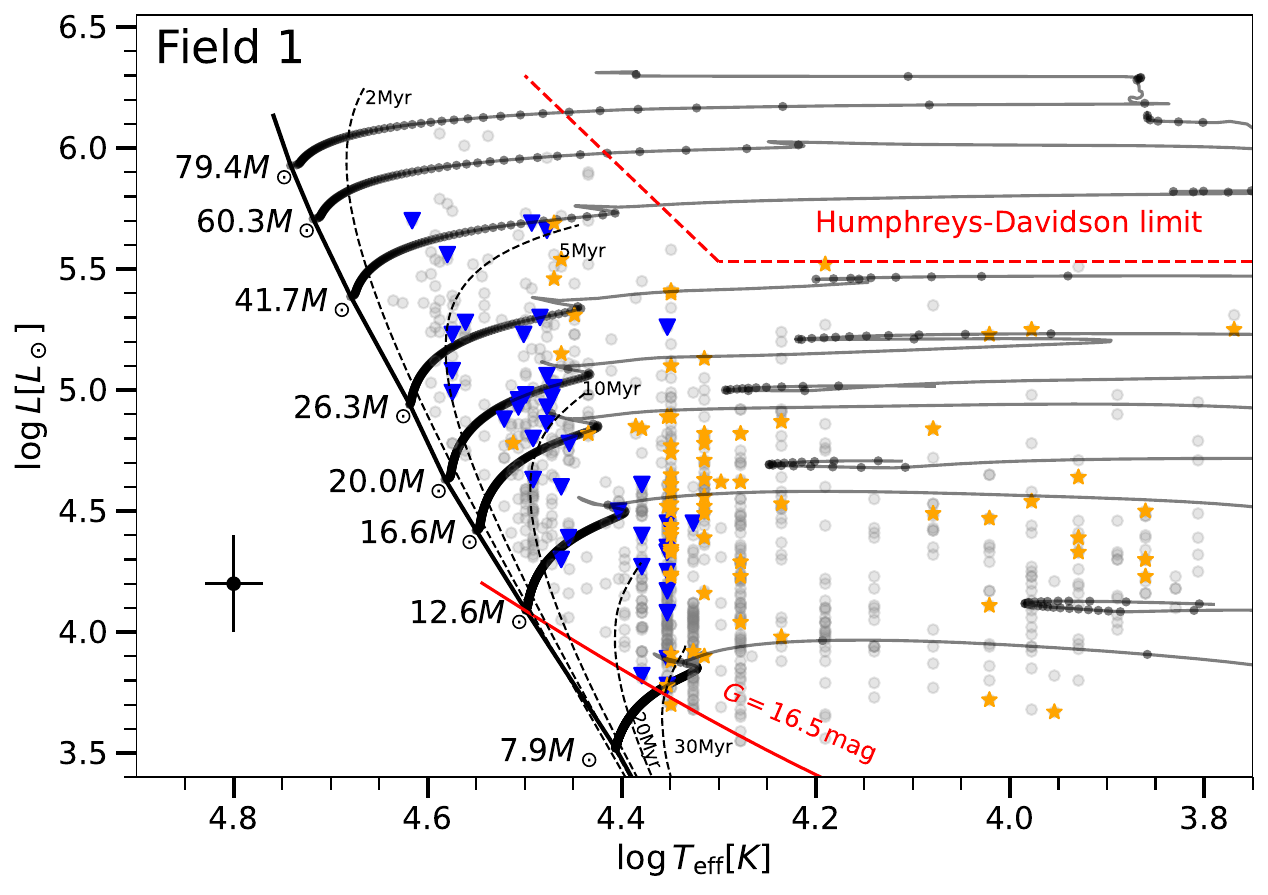} & 
\includegraphics[width=0.45\textwidth]{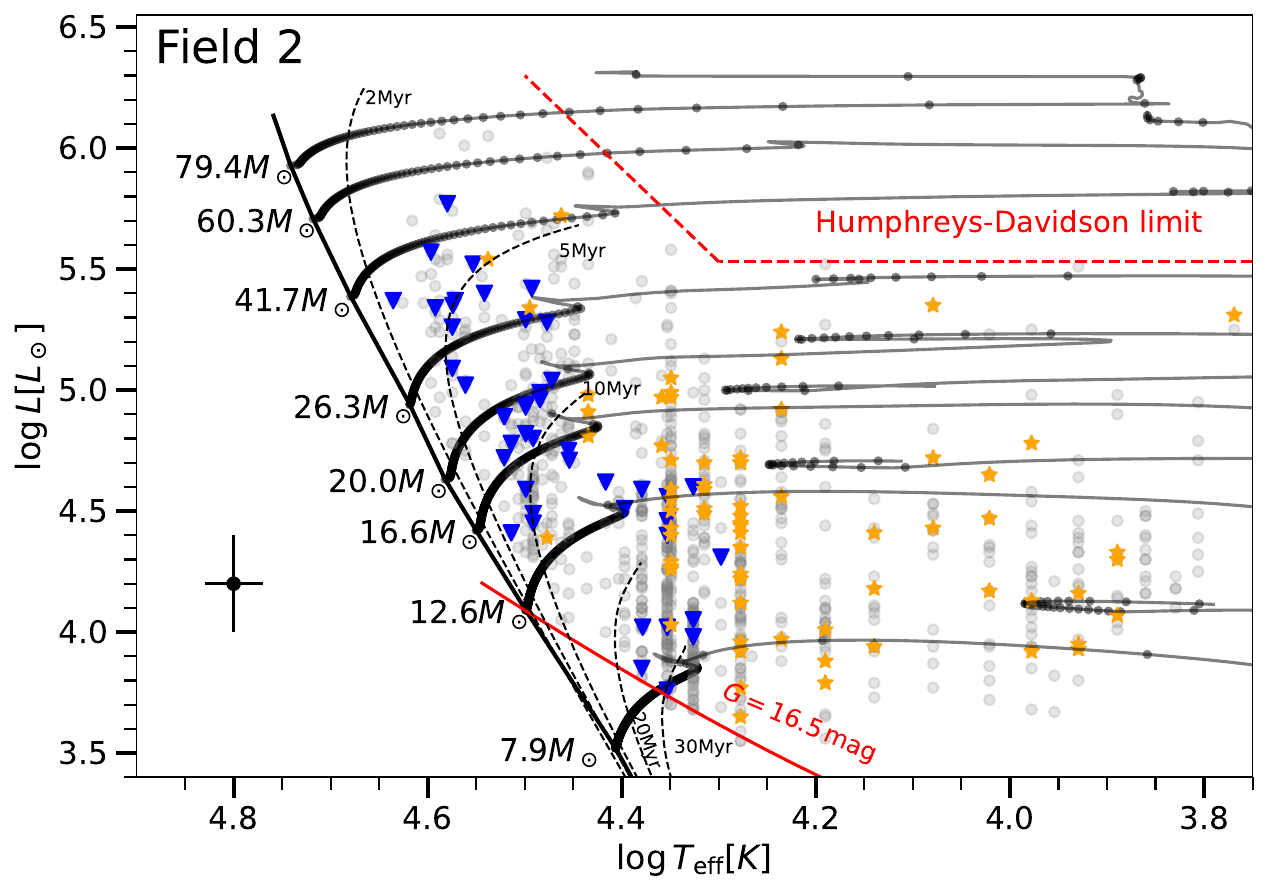} \\ 
\includegraphics[width=0.45\textwidth]{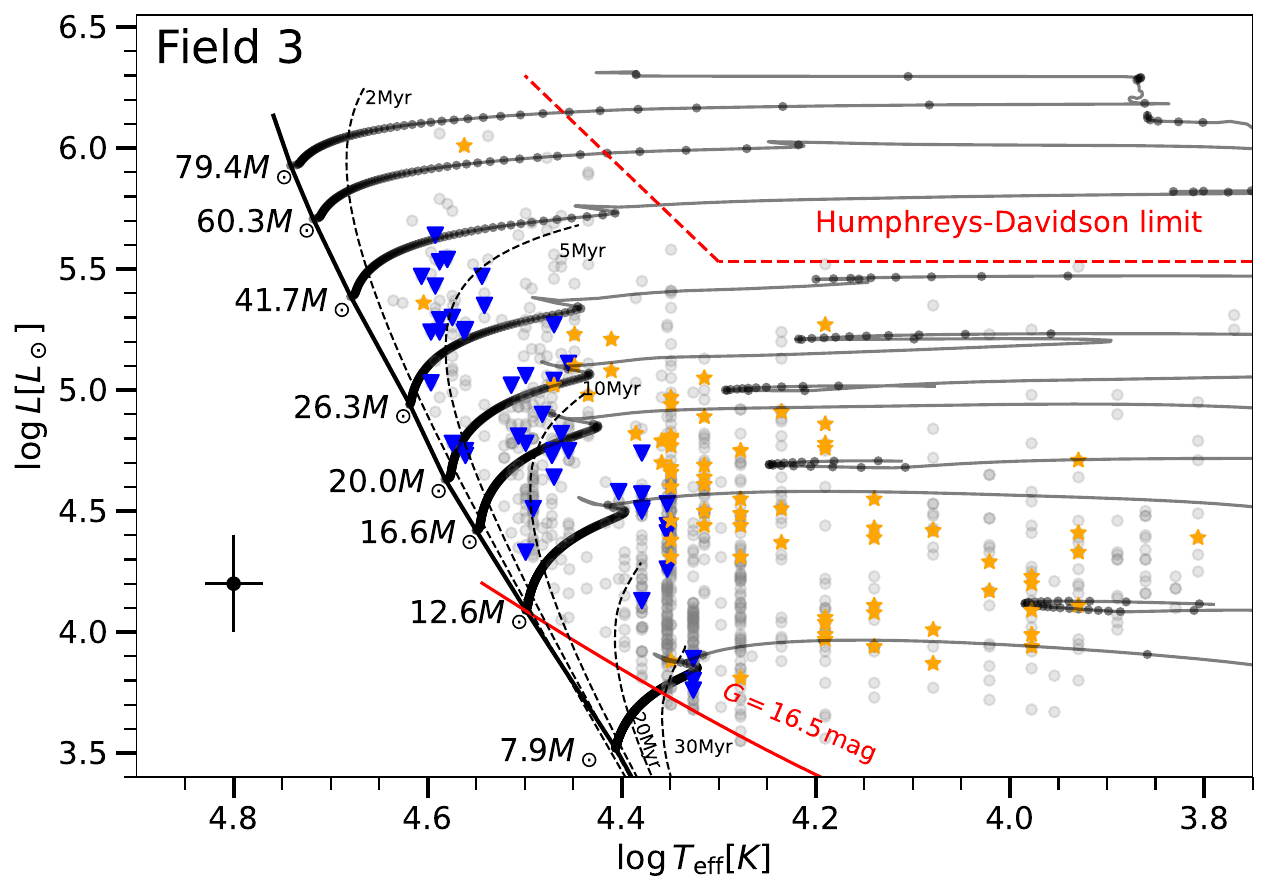} & 
\includegraphics[width=0.45\textwidth]{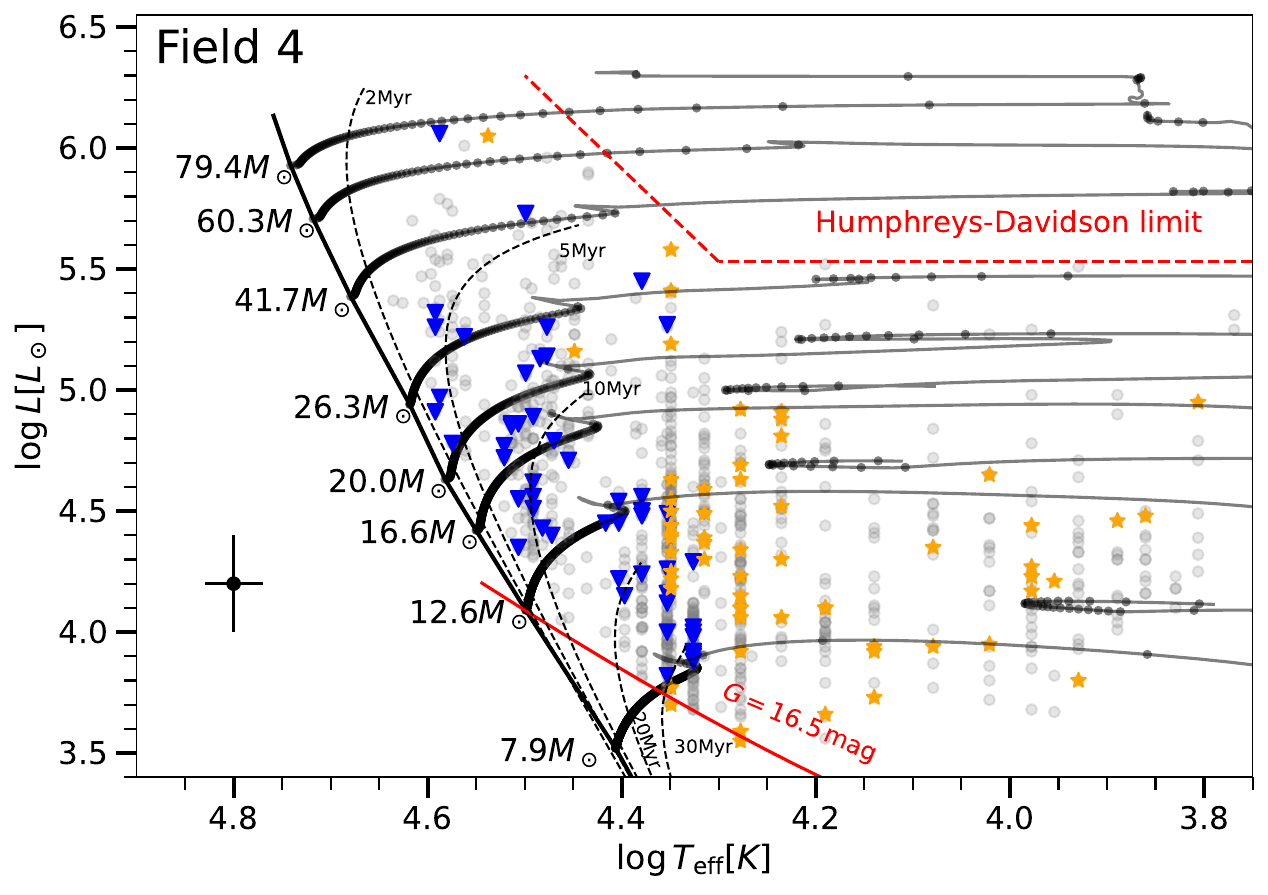} \\ 
\includegraphics[width=0.45\textwidth]{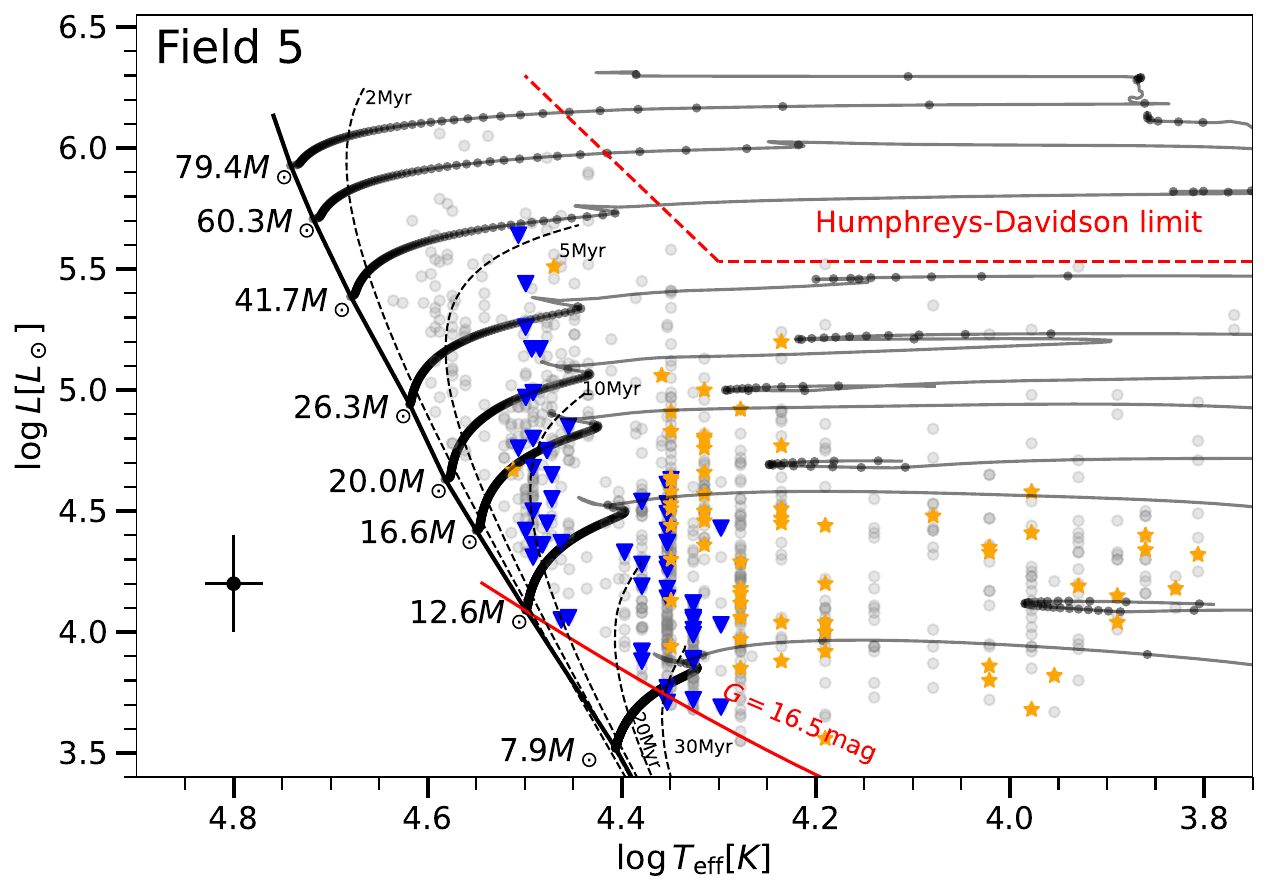} & 
\includegraphics[width=0.45\textwidth]{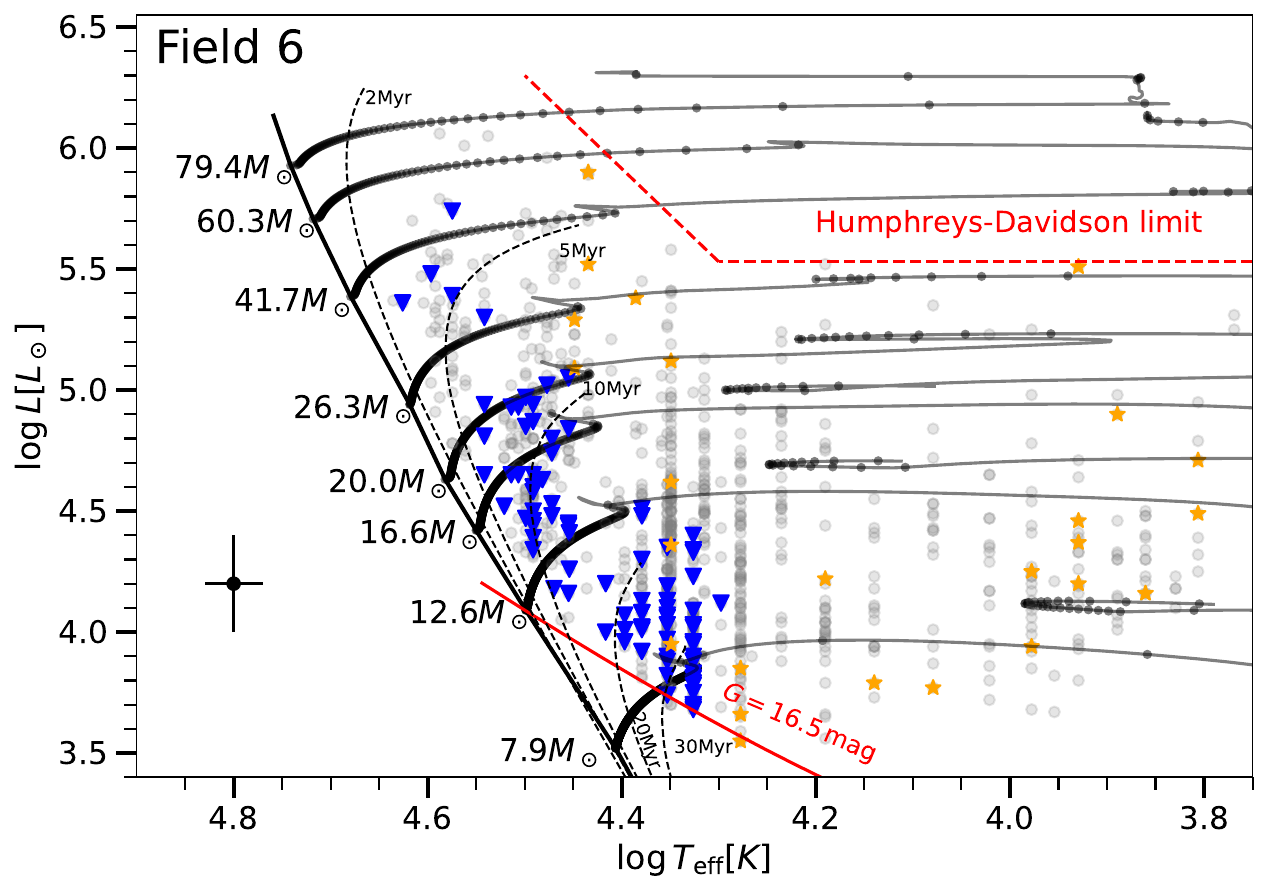} \\ 
\includegraphics[width=0.45\textwidth]{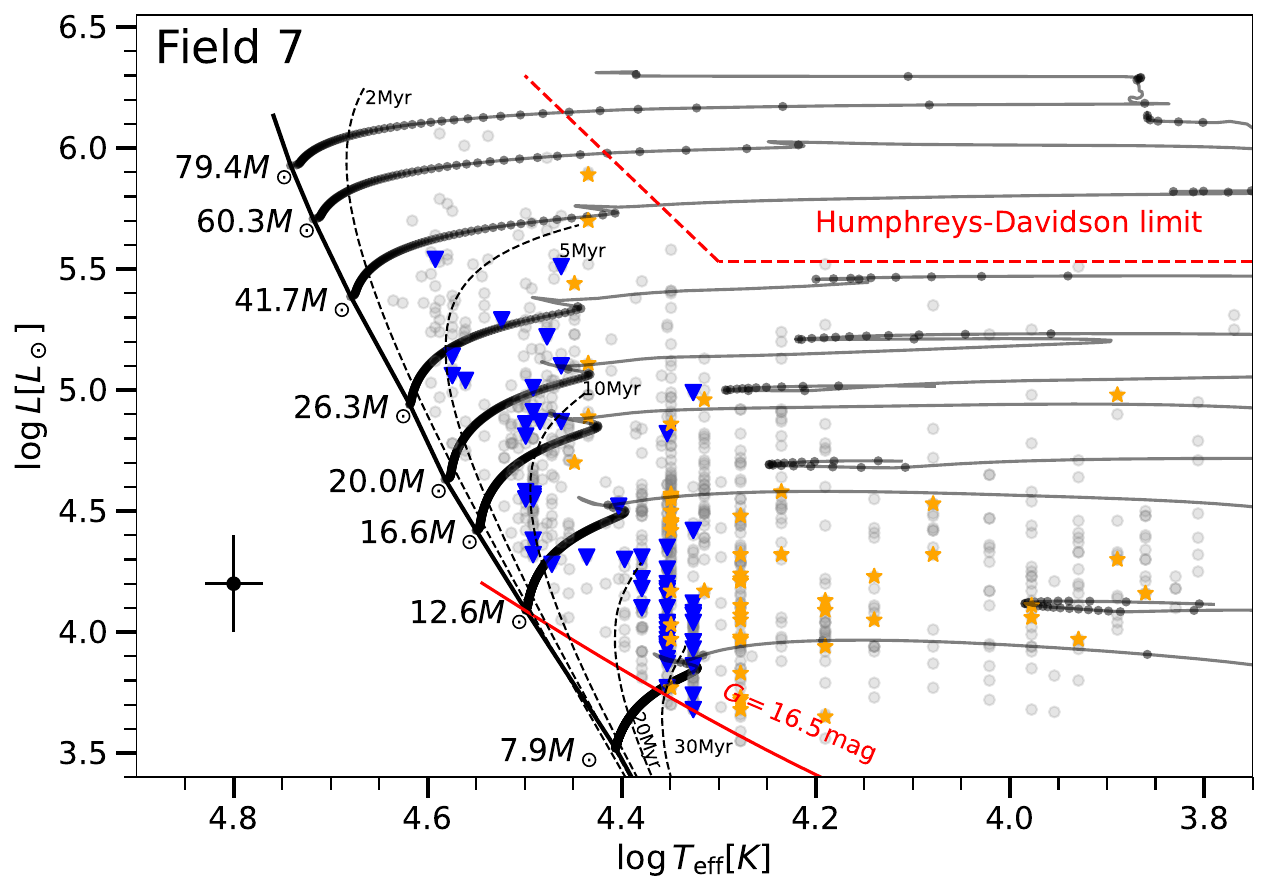} & 
\includegraphics[width=0.45\textwidth]{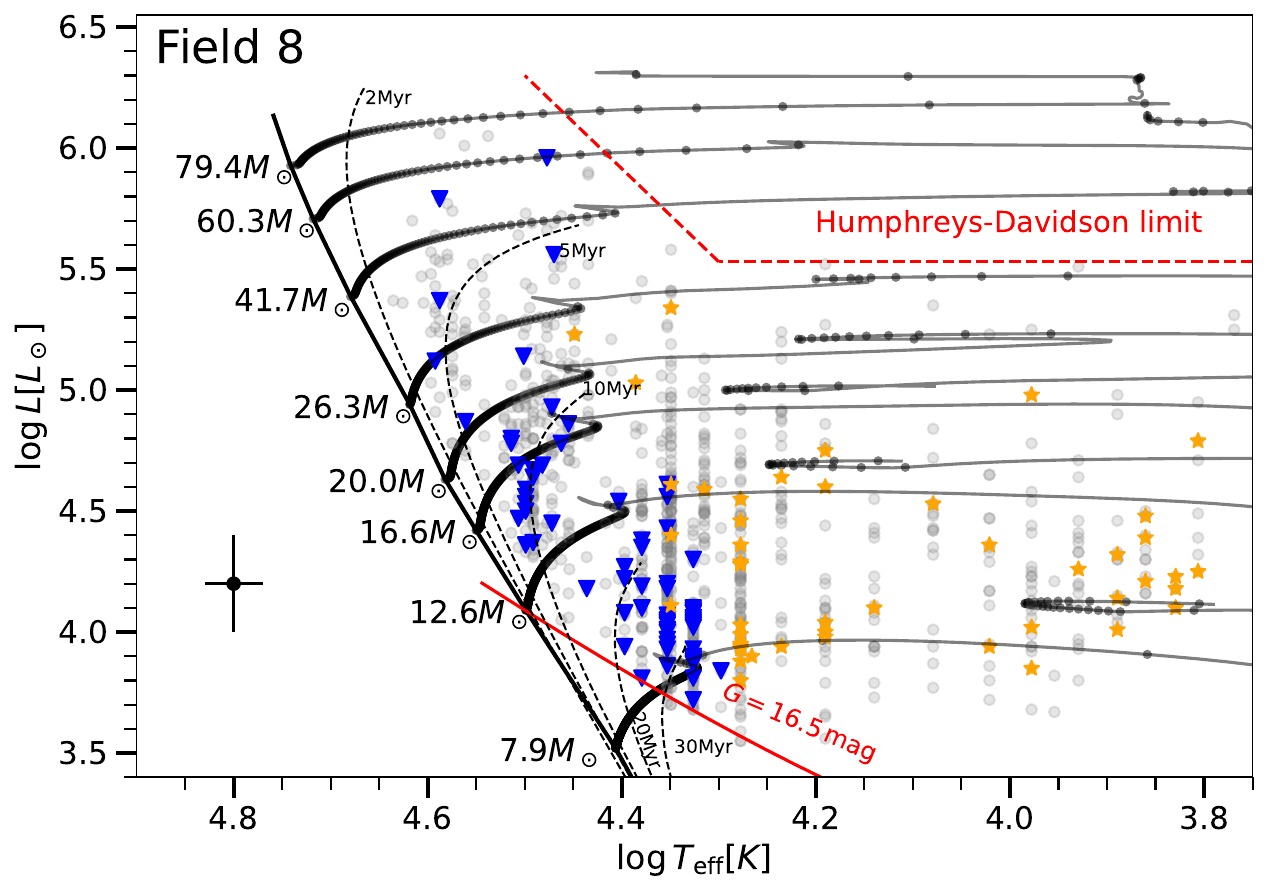} \\ 
\end{tabular}
\caption{Same as Fig.\,\ref{fig:HRD}, but with the samples of the eight SMC fields shown in Fig.\,\ref{fig:BLOeMOverview} highlighted in colour (colour meaning is the same as in Fig.\,\ref{fig:HRD}). The entire sample is shown in grey in each panel. }
\label{fig:FieldHRD}
\end{figure*}

Figure\,\ref{fig:FieldHRD} shows the HRDs of each of the eight SMC fields observed in the framework of BLOeM (shown in Fig.\,\ref{fig:BLOeMOverview}). The differences between the populations are not blatant. Generally, fields 1 -- 4 contain a higher number of stars "born as O-type" ($M_{\rm ini} \gtrsim 14\,M_\odot$) compared to fields 5 -- 8, which could be anticipated given the automated way with which the fields were selected, prioritizing fields with the most massive stars (Sect.\,\ref{sec:sample}). Field 5 appears to contain the oldest population among the eight fields.

\end{appendix}

\end{document}